\documentclass[12pt,urlcolor=black, linkcolor=black]{article}
\usepackage{savesym}
\usepackage{cite}
\usepackage{wasysym}
\usepackage{amscd}
\usepackage{graphicx}
\usepackage{pifont}
\usepackage{float}
\usepackage{subfig}
\usepackage[all]{xy}
\usepackage{multicol}
\usepackage{amsfonts}
\usepackage{picture}
\usepackage{amssymb}
\savesymbol{iint}
\savesymbol{iiint}
\usepackage{amsmath}
\usepackage{amsthm}
\restoresymbol{TXF}{iint}
\restoresymbol{TXF}{iiint}
\usepackage{PU}
\usepackage{color}
\usepackage[colorlinks,citecolor=black]{hyperref}
\hypersetup{colorlinks=true}
\hypersetup{linkcolor=black}
\hypersetup{citecolor=black}
\hypersetup{urlcolor=black}
\usepackage{setspace}
\usepackage{multirow}
\usepackage{wrapfig}
\usepackage{verbatim}
\usepackage{dsfont}
\usepackage{youngtab}
\numberwithin{equation}{section}
\usepackage{float}
\restylefloat{table}
\allowdisplaybreaks
\usepackage{accents}

\usepackage{slashed}
\usepackage{bm}

\newcommand{\CPoT}{\CC\CP_{\perp}\T}

\usepackage{mathtools}

\def\tilde{\widetilde}

\def\bar{\overline}

\def\1{{\mathds 1}}

\def\CC{{\mathcal C}}

\def\CL{{\mathcal L}}
\def\CM{{\mathcal M}}

\def\CP{{\mathcal P}}

\def\CR{{\mathcal R}}

\def\CU{{\mathcal U}}

\newcommand{\Z}{\mathbb{Z}}
\newcommand{\Tr}{\mathrm{Tr}}
\newcommand{\T}{\mathcal{T}}

\newcommand{\Sq}{\mathrm{Sq}}

\usepackage{datetime}
\usepackage{dsfont}

\def\Sq{\mathrm{Sq}}

\newcommand{\SO}{{\rm SO}}
\newcommand{\Spin}{{\rm Spin}}
\newcommand{\Sp}{{\rm Sp}}
\newcommand{\U}{{\rm U}}

\newcommand{\SU}{{\rm SU}}
\newcommand{\PSU}{{\rm PSU}}

\newcommand{\ii}{\hspace{1pt}\mathrm{i}\hspace{1pt}}
\newcommand{\dd}{\hspace{1pt}\mathrm{d}}

\newcommand{\rH}{{\rm H}}

\def\bZ{{\mathbf{Z}}}

\newcommand{\bea}{\begin{eqnarray}}
\newcommand{\eea}{\end{eqnarray}}
\def\be{\begin{equation}}
\def\ee{\end{equation}}

\newcommand{\Wmfootnote}[1]{%
\let\oldthefootnote=\thefootnote%
\stepcounter{mpfootnote}%
\addtocounter{footnote}{-1}%
\renewcommand{\thefootnote}{{$W^-$}}
\footnote{#1}%
\let\thefootnote=\oldthefootnote%
}

\newcommand{\Wpfootnote}[1]{%
\let\oldthefootnote=\thefootnote%
\stepcounter{mpfootnote}%
\addtocounter{footnote}{-1}%
\renewcommand{\thefootnote}{{{W$^+$\,}}}
\footnote{#1}%
\let\thefootnote=\oldthefootnote%
}

\newcommand{\Yfootnote}[1]{%
\let\oldthefootnote=\thefootnote%
\stepcounter{mpfootnote}%
\addtocounter{footnote}{-1}%
\renewcommand{\thefootnote}{{{Y$^-$\;}}}
\footnote{#1}%
\let\thefootnote=\oldthefootnote%
}

\newcommand{\Zfootnote}[1]{%
\let\oldthefootnote=\thefootnote%
\stepcounter{mpfootnote}%
\addtocounter{footnote}{-1}%
\renewcommand{\thefootnote}{{{Z$^0$\;\;}}}
\footnote{#1}%
\let\thefootnote=\oldthefootnote%
}

\begin{document}

\title{Gauge Enhanced Quantum Criticality and 
Time Reversal Domain Wall: 
 $\SU(2)$ Yang-Mills Dynamics with Topological Terms}

\authors{Juven Wang \worksat{\CMSA, \IAS}\Wpfootnote{e-mail: {\tt jw@cmsa.fas.harvard.edu}}, Yi-Zhuang You \worksat{\UCSD}\Yfootnote{e-mail: {\tt yzyou@physics.ucsd.edu}}, Yunqin Zheng \worksat{\Princeton}\Zfootnote{e-mail: {\tt yunqinz@princeton.edu}}}

\institution{CMSA}{\centerline{${}^{1}$Center of Mathematical Sciences and Applications, Harvard University, Cambridge, MA 02138, USA}}
\institution{IAS}{\centerline{${}^{2}$School of Natural Sciences, Institute for Advanced Study, Princeton, NJ 08540, USA}}
\institution{UCSD}{\centerline{${}^{3}$Department of Physics, University of California, San Diego, CA 92093, USA}}
\institution{Princeton}{\centerline{${}^{4}$Physics Department, Princeton University, Princeton, NJ 08540, USA}}

\abstract{

We explore the low energy dynamics of the four siblings of Lorentz symmetry enriched $\SU(2)$ Yang-Mills theory with a theta term at $\theta=\pi$ in $(3+1)$d.  Due to a mixed anomaly between time reversal symmetry and the center symmetry, the low energy dynamics must be nontrivial. We focus on two possible scenarios: 1) time reversal symmetry is spontaneously broken by the two confining vacua, and 2) the  low energy theory is described by a $\U(1)$ Maxwell gauge theory (e.g. $\U(1)$ spin liquid in condensed matter) which is deconfined and gapless while preserving time reversal symmetry. In the first scenario, we first identify the global symmetry on the time reversal domain wall, where time reversal symmetry in the bulk induces a $\Z_2$ unitary symmetry on the domain wall. We discuss how the Lorentz symmetry and the unitary $\Z_2$ symmetry enrich the domain wall theory.    In the second scenario, we relate the symmetry enrichments of the $\SU(2)$ Yang-Mills to that of the $\U(1)$ Maxwell gauge theory. 
This further opens up the possibility that $\SU(2)$ QCD with large and odd flavors of fermions could be a direct second order phase transition between two phases of $\U(1)$ gauge theories as well as between a $\U(1)$ gauge theory and a trivial vacuum (e.g. a trivial paramagnet), where the gauge group is enhanced to be non-Abelian at and only at the transition.
We characterize these transitions, and name them as \emph{Gauge Enhanced Quantum Critical Points}.

}

\maketitle

\pagenumbering{arabic}
\setcounter{page}{2}

\setcounter{tocdepth}{3}
\tableofcontents

\clearpage

\section{Introduction}

The $\SU(N)$ Yang-Mills theory is a non-Abelian gauge theory with a gauge group $\SU(N)$ described by the action
\begin{eqnarray}
S= -\frac{1}{4g^2} \int_{M_4} \Tr(F \wedge \star F) + \frac{\theta}{8\pi^2} \int_{M_4} \Tr(F\wedge F),
\end{eqnarray}
which admits a topological term parameterized by a variable $\theta$.   Since the second Chern number
\begin{eqnarray}
c_2(V_{\SU(N)})= \frac{1}{8\pi^2}  \Tr(F\wedge F)
\end{eqnarray}
of the $\SU(N)$ vector bundle integrates to be an integer, $\theta$ is $2\pi$ periodic\cite{Dashen:1970et, 2017JHEP...05..091G}. The theory has a $\Z_{2,[1]}$ one form center symmetry\cite{Kapustin:2013qsa, Kapustin:2013uxa, Kapustin2014gua1401.0740, Gaiotto:2014kfa}. When $\theta=0, \pi\mod 2\pi$, it is also time reversal symmetric.

$\SU(N)$ Yang-Mills is the simplest non-Abelian gauge theory in $3+1$d that exhibits rich dynamics. In contrast to the Abelian $\U(1)$ Maxwell gauge theory which is free, the $\SU(N)$ Yang-Mills is strongly coupled due to negative beta function, and the low energy dynamics is prohibitive via merely perturbative approaches\cite{YMMP-Jaffe-Witten}. However, various evidences including 't Hooft anomalies\cite{tHooft:1980xss, 2017JHEP...05..091G}, deformation of supersymmetric Yang-Mills\cite{Evans:1996hi, Konishi:1996iz, AharonyASY2013hda1305.0318}, and holographic calculation in the large $N$ limit\cite{Witten:1998uka, Vicari:2008jw} provide various constraints on the low energy dynamics, which we summarize as the \emph{Standard Lore} of Yang-Mills. 

\subsection{Standard Lore of $\SU(N)$ Yang-Mills}
\label{sec. lore}

We review the dynamics of $\SU(N)$ Yang-Mills as a function of $\theta\in [0, 2\pi)$. 
\begin{itemize}
	\item $\theta=0$: When $\theta=0$, the only term is the kinetic energy of the gauge field, which is time reversal symmetric. Various evidences including lattice simulations, softly broken supersymmetry and large $N$ holographic models suggest that the ground state is confining with an unbroken center symmetry,  and there is a mass gap\cite{YMMP-Jaffe-Witten, 2017JHEP...05..091G, Evans:1996hi, Konishi:1996iz, AharonyASY2013hda1305.0318}. 
	\item $\theta=\pi$: Another instance which is time reversal symmetric is when $\theta=\pi$. In this case, there is a mixed anomaly between the time reversal symmetry and the $\Z_N$ center symmetry for even $N$, and a more subtle global inconsistency for odd $N$\cite{2017JHEP...05..091G, WanWangZheng190400994, Wan2018zql1812.11968}. Both cases are unified from the point of view of anomaly in the space of coupling constants\cite{Cordova:2019jnf, Cordova:2019uob}.  For even $N$, this anomaly immediately constrains that $\SU(N)$ Yang-Mills with $\theta=\pi$ can not flow to a trivial phase. It is widely believed that at the low energy, the theory confines and the center symmetry is unbroken. However time reversal is spontaneously broken, leading to two degenerate ground states\cite{Dashen:1970et, PhysRevLett.92.201601, 2017JHEP...05..091G}. Such spontaneous broken of time reversal has been shown for large $N$ Yang-Mills, where as one tunes from $\theta<\pi$ to $\theta>\pi$ a first order phase transition has been observed\cite{Witten:1998uka, Vicari:2008jw}. 
	\item $0<\theta<\pi, \pi<\theta<2\pi$: The dynamics in this regime is less clear, due to the lack of time reversal symmetry and consequently the anomaly.  It is believed that the theory confines for all $\theta$. This is also supported by the large $N$ calculation\cite{Witten:1998uka, Vicari:2008jw}. The phase at $\theta=2\pi$, although is believed to be dynamically trivial as $\theta=0$, differs from the phase at $\theta=0$  by a subtle symmetry protected topological (SPT) phase \cite{Kapustin:2013qsa, Kapustin:2013uxa, AharonyASY2013hda1305.0318, Gaiotto:2014kfa, Hsin2018vcg1812.04716}.
\end{itemize}

Although the Standard Lore is believed to hold for large $N$,  there are less evidences supporting the standard lore for small $N$. In particular, for $N=2$, the $\SU(2)$ Yang-Mills at $\theta=\pi$ can flow to one of the several possible scenarios at low energy.  The low energy theory should either spontaneously break time reversal, or be deconfined, or preserve time reversal symmetry and being confined while being gapless. 
\footnote{Gapped and confined TQFT that preserve time reversal symmetry has been ruled out in a recent work\cite{COnogoTQFT}. In \cite{WanWangZheng190400994, Wan2018zql1812.11968}, the authors constructed a $H$-symmetry extended TQFT via the exact sequence $1\to K\to H\to \Z_{2,[1]}\to 1$,  generalizing \cite{Wang2017locWWW1705.06728, 2018PTEP1801.05416} to higher form symmetries. By dynamically gauging $K$, it has been realized that $\Z_{2,[1]}$ is spontaneously broken, which is consistent with \cite{COnogoTQFT}. }
As far as we know, none of the above scenarios has been excluded for $N=2$.  Therefore, it is desirable to study all possible scenarios of $\SU(2)$ Yang-Mills in detail.

\subsection{New Aspects: Lorentz Symmetry Enrichments}
\label{sec.newaspect}

For any gauge theory with $\Z_{2,[1]}$ one form symmetry, and in particular the $\SU(2)$  Yang-Mills with any theta parameter\cite{WanWangZheng190400994, Wan2018zql1812.11968}, can be enriched by the $\SO(3,1)$ Lorentz symmetry,\footnote{There are two branches of $\SO(3,1)$, differed by chirality. These are denoted as $\SO^{\pm}(3,1)$ in the literature. For our purposes, the choice of chirality will not play a role. In the rest of the paper, we focus on the positive chirality $+$, and will suppress the superscript for simplicity.  } via fractionalizing the Lorentz symmetry on the Wilson line operators. This phenomena has been previously explored in \cite{Witten:1988ze, Guo:2017xex, Cordova2018acb1806.09592} and others, and has been recently termed in \cite{2019arXiv190907383H} poetically as Lorentz symmetry fractionalization.
Fractionalization of the Lorentz symmetry on a Wilson line requires that the Wilson line transforms projectively under $\SO(3,1)$, i.e., the self statistics is shifted by $h=1/2$. This is done by shifting the background field $B$ for the center $\Z_{2,[1]}$ one-form symmetry by the second Stiefel-Whitney class of the tangent bundle of the spacetime manifold, i.e., 
\begin{eqnarray}
B\to B+ K_2 w_2,
\end{eqnarray}
where $K_2=0,1$ represents trivial/nontrivial fractionalization.

For $\theta=0, \pi$, the $\SU(2)$ Yang-Mills is time reversal symmetric. Thus one can further enrich the $\SU(2)$ Yang-Mills by the time reversal symmetry (or $\mathrm{O}(3,1)$ if combined with the $\SO(3,1)$ Lorentz symmetry)\cite{WanWangZheng190400994, Wan2018zql1812.11968}.  In this case,  time reversal symmetry can be fractionalized on the Wilson line. Nontrivial fractionalization of time reversal means that the Wilson line is a Kramers doublet. Formally, this is done by shifting the background field $B$ by the square of the first Stiefel-Whitney class of the tangent bundle of the spacetime manifold, i.e.
\begin{eqnarray}
B\to B+ K_1 w_1^2,
\end{eqnarray}
where $K_1=0,1$ represents trivial/nontrivial fractionalization of time reversal symmetry. Of course, one can consider enriching the $\SU(2)$ Yang-Mills at $\theta=0,\pi$ by both time reversal and $\SO(3,1)$ Lorentz symmetry.  
In \cite{WanWangZheng190400994, Wan2018zql1812.11968}, we denote the four different $\mathrm{O}(3,1)$ Lorentz symmetry enrichments, labeled by $(K_1, K_2)$,  of $\SU(2)$ Yang-Mills at $\theta=0,\pi$  as the \emph{Four Siblings}, which we use throughout the present work.

Keeping the symmetry enrichments in mind, it is natural to revisit the standard lore and ask a more refined question: \textbf{How the dynamics of $\SU(N)$ Yang-Mills depends on the $\mathrm{O}(3,1)$ symmetry enrichment, i.e. the four siblings $(K_1, K_2)$? }
In this work, we study the dynamics of $\SU(2)$ Yang-Mills at $\theta=\pi$, and focus on two low energy scenarios. 

In the first scenario (to be discussed in section \ref{Sec. TimeRevDW}), time reversal is spontaneously broken, and we study the domain wall theory that is constrained by the 't Hooft anomalies. We highlight several features of our results: 
\begin{enumerate}
	\item The domain wall theory is not time reversal symmetric, in contrast to the bulk. Instead, there is a discrete unitary symmetry $\CU$. 
	\item  The four siblings in the bulk corresponds to four different enrichments of the Lorentz symmetry as well as the unitary symmetry $\CU$ on the wall. 
	\item Even though the 't Hooft anomaly of  $\SU(2)$ Yang-Mills does not depend on the $\SO(3,1)$ Lorentz symmetry enrichment, the 't Hooft anomaly on the wall does.  
\end{enumerate}
In section \ref{sec.app1}, we also discuss the consequences of symmetry enrichments on the domain wall theories for $\SU(2)$ QCD within the regime of chiral symmetry breaking $N_f<N_{CFT}$. 

The second scenario will be discussed in section \ref{Sec.U1}, where we assume that the low energy of $\SU(2)$ Yang-Mills with $\theta=\pi$ is deconfined. In particular, we only discuss the case where the low energy theory is described by a $\U(1)$ Maxwell theory, with certain Lorentz symmetry enrichment. The Lorentz symmetry enriched $\U(1)$ Maxwell theories have been studied in \cite{Wang2016cto1505.03520, Zou2017ppq1710.00743,  Hsin2019fhf1904.11550, NingZouMeng2019} where they classify the phases of time reversal $\U(1)$ quantum spin liquids.  We will use the 't Hooft anomaly to constrain the correspondence between the symmetry enrichments of $\SU(2)$ Yang-Mills at $\theta=\pi$ and the symmetry enrichments of the Maxwell theory. In section \ref{Sec.app2}, we further apply this correspondence to study the phase transitions between different $\U(1)$ spin liquids, as well as the phase transitions between $\U(1)$ spin liquids and trivial paramagnets. Amusingly, we find that $\SU(2)$ QCD with $N_f$ fermions ($N_f>N_{CFT}$) in the fundamental representation can be interpreted as the second order phase transition between the the above phases, where the gauge group is enhanced at and only at the transition point. We denote such transition as \emph{gauge enhanced quantum critical points}.

\section{SU(2) Yang-Mills Theory at $\theta=\pi$}
\label{sec.SU2YM}

\subsection{Four Siblings and Anomalies}
\label{sec.YM}

We start by reviewing the results in \cite{WanWangZheng190400994, Wan2018zql1812.11968}. The 4d SU(2) Yang-Mills gauge theory with an SU(2) gauge group and
a theta term in the Minkowski spacetime $M_4$ is described by an action \footnote{For definiteness, the spacetime signature is taken to be $(-1, 1, 1, 1)$. } 
\begin{eqnarray}\label{SU2YMaction}
S= -\frac{1}{4g^2} \int_{M_4} \Tr(F \wedge \star F) + \frac{\theta}{8\pi^2} \int_{M_4} \Tr(F\wedge F),
\end{eqnarray}
where we denote $a$ as the SU(2) gauge field and $F=\dd a-\ii a^2$ is the field strength. 
The partition function is 
\bea \label{ZYM}
\bZ^{\text{$4$d}}_{{\text{YM}}}
\equiv\int [{\cal D} {a}] \exp( \ii S ).
\eea
Since the second Chern number $c_2(V_{\SU(2)})= \int_{M_4}\Tr(F\wedge F)/8\pi^2$ is quantized to be an integer, the $\theta$ parameter has periodicity $2\pi$.  \eqref{SU2YMaction} is time reversal symmetric only when $\theta=0, \pi\mod 2\pi$. The readers may refer to \cite{WanWangZheng190400994} for details of time reversal transformations on the gauge fields, as well as various peculiarities about $\SU(2)$.

The theory also has a $\Z_{2,[1]}$ one-form center symmetry that acts by shifting the connection by a $\Z_2$ flat connection. We can turn on a background $\Z_2$ two-form gauge field $B\in \rH^2(M_4, \Z_2)$ for this one-form symmetry.\footnote{For simplicity, we will also use the same symbol $B$ as a representative in $\rH^2(M_4, \Z_2)$, i.e., $B$ is a cocycle satisfying the cocycle condition $\delta B= 0\mod 2$. } In the presence of this background gauge field, the SU(2) bundle is twisted into a $\PSU(2)\equiv \SO(3)$ bundle with fixed second Stiefel-Whitney class 
\begin{eqnarray}\label{GBC1}
w_2(V_{\SO(3)})= B.
\end{eqnarray}
To see how $B$ couples to the theory \eqref{SU2YMaction}, it is convenient to promote the SU(2) gauge field $a$ to a U(2) gauge field $\widehat{a}$\cite{2017JHEP...05..091G}. It is instructive to realize that $w_2(V_{\SO(3)}) = c_2(V_{\U(2)}) \mod 2$. Hence \eqref{GBC1} can be equivalently written as
\begin{eqnarray}\label{GBC2}
c_2(V_{\U(2)}) = B \mod 2.
\end{eqnarray} 
As we are focusing on the time reversal symmetric theory, one should be tempted to formulate the theory \eqref{SU2YMaction} on an unorientable manifold. The Lorentz symmetry associated with an unorientable manifold is O(3,1).  In particular, on a generic unorientable manifold, both the first and second Stiefel-Whitney classes, $w_1$ and $w_2$, of the tangent bundle of the spacetime manifold $M_4$  are allowed to be nontrivial. One can twist the gauge bundle constraint \eqref{GBC2} as 
\begin{eqnarray}\label{GBC3}
c_2(V_{\U(2)}) = B+ K_1 w_1^2+ K_2 w_2 \mod 2, ~~~~~ K_1, K_2=0,1.
\end{eqnarray}
As explained in section \ref{sec.newaspect}, $(K_1, K_2)$ labels four distinct $\mathrm{O}(3,1)$ Lorentz symmetry enrichments of $\SU(2)$ Yang-Mills theories. In \cite{WanWangZheng190400994, Wan2018zql1812.11968}, the authors  also referred $(K_1, K_2)$ as the \emph{Four Siblings} of $\mathrm{O}(3,1)$  enriched  SU(2) Yang-Mills with $\theta=0, \pi$. 

One can understand \eqref{GBC3} as follows. We start with \eqref{GBC1}. When $B$ is nontrivial, the Wilson line with SU(2) isospin $j=1/2$, $W_{1/2}(\gamma)$, is attached to a surface operator $\exp(\ii \pi \int_\Sigma B)$ with $\partial \Sigma=\gamma$. The twisted gauge bundle constraint modifies the above surface operator by decorating an additional 2d invertible TQFT of the Lorentz symmetry: $\pi(K_1 w_1^2+ K_2 w_2)$. The physical meanings of these invertible TQFTs are well known.  $\pi w_1^2$ is the worldsheet theory of a time reversal symmetric SPT (a.k.a. the Haldane chain) whose boundary supports a Kramers doublet. $\pi w_2$ is the worldsheet theory whose boundary transforms projectively under the Lorentz symmetry SO(3,1), i.e., the boundary supports a fermion. We further realize that without the twists from the Lorentz symmetry O(3,1) (i.e. $K_1=K_2=0$),  the original SU(2) Wilson line $W_{1/2}(\gamma)$ transforms under O(3,1) as Kramers singlet and is a boson.  Combining the above physical understandings, under the twists using the O(3,1) Lorentz symmetry, the statistics $h(W)$ and the Kramers parity $\T_{W}^2$ of $W_{1/2}(\gamma)$ are
\begin{eqnarray}
h(W)= \frac{K_2}{2}\mod 1, ~~~~~ \T_{W}^2= (-1)^{K_1+K_2}.
\end{eqnarray}
It is also illuminating to refer twisting the gauge bundle constraint from \eqref{GBC2} to \eqref{GBC3} as Lorentz symmetry fractionalization. See \cite{2019arXiv190907383H} for the related discussions on 3d Chern-Simons (matter) theories and \cite{Hsin2019fhf1904.11550, NingZouMeng2019} on 4d U(1) gauge theories.

The SU(2) Yang-Mills theory with $\theta=\pi$, coupled to the two-form background field $B$, is
\begin{equation}\label{SU2actionB}
S= -\frac{1}{4g^2} \int_{M_4} \Tr(\widehat{F}-\pi B I_2) \wedge \star (\widehat{F}-\pi B I_2) + \frac{\pi}{8\pi^2} \int_{M_4} \Tr(\widehat{F}-\pi B I_2)\wedge(\widehat{F}-\pi B I_2),
\end{equation}
subjected to the gauge bundle constraint \eqref{GBC3}. Here $\widehat{F}= \dd\widehat{a}-\ii \widehat{a}^2$ is the $\U(2)$ field strength.  One further attempts to formulate \eqref{SU2actionB} on an unorientable and non-spin manifold $M_4$, which enables one to prove the \emph{full} quantum anomalies.  It was found in \cite{WanWangZheng190400994} that when $B$ is nontrivial, \eqref{SU2actionB} can not be consistently defined on an unorientable $M_4$. The resolution is to define \eqref{SU2actionB} as a 4d-5d coupled system, where the structures $(w_1, w_2, B)$ on $M_4$ are extended to $M_5$.\footnote{For simplicity, we denote the first and second Stiefel-Whitney classes of \emph{both} $M_4$ and $M_5$ as  $w_1$ and $w_2$. } In particular, $M_5$ is unorientable. The Anomaly polynomial is
\begin{eqnarray}\label{anomSU2}
S_{\mathrm{anom}}= \pi\int_{M_5} B \Sq^1 B + \Sq^2 \Sq^1 B + K_1 w_1^2 \Sq^1 B + K_2 \Sq^1(w_2 B).
\end{eqnarray}

We emphasize that the anomaly polynomial \eqref{anomSU2} depends on $K_2$ only when $M_5$ has a nontrivial boundary $M_4$. This implies that the term $K_2 \Sq^1(w_2 B)$ does not lead to a distinguished anomaly. Instead, it is a WZW-like counter term. However, we will show in section \ref{Sec. TimeRevDW} that, if time reversal is spontaneously broken at $\theta=\pi$,  the WZW-like counter term leads to a nontrivial 't Hooft anomaly on the time reversal domain wall.

\subsection{Low Energy Dynamics: Overview and Questions}

The SU(2) Yang-Mills theory is strongly coupled in the infrared, due to negative beta function. Thus the low energy fate of the SU(2) dynamics is hardly known. 
It is famously conjectured \cite{YMMP-Jaffe-Witten} that for any $N\geq 2$ the $\SU(N)$ Yang-Mills with $\theta=0$ has a mass gap.\footnote{Though the mass gap is supported by numerous evidences, it still remains a conjecture. In section \ref{Sec.app2}, we contemplate another exotic possibility where the low energy of $\theta=0$ Yang-Mills is gapless, described by a deconfined $\U(1)$ Maxwell theory.}   Moreover, \cite{2017JHEP...05..091G}  found that the $\SU(N)$  Yang-Mills (for even $N$) has a nontrivial 't Hooft anomaly only at $\theta=\pi$. 
Since nontrivial 't Hooft anomaly implies that the low energy theory can not be trivially gapped, 
there should be nontrivial dynamics at $\theta=\pi$. In particular, the above analysis also apply to SU(2) Yang-Mills.

For the regime within $\theta\in (0, \pi)\cup (\pi, 2\pi)$, the dynamics is less clear. In fact, \cite{2017JHEP...05..091G} proposed two scenarios 
for the SU(2) Yang-Mills dynamics at zero temperature. In one scenario,  SU(2) Yang-Mills is confined for every $\theta$. 
In the other scenario, SU(2) Yang-Mills is deconfined within a regime $\theta\in [\pi-x, \pi+x]$ for $x\in [0, \pi)$. In the following discussion, we will not discuss the generic $\theta$ and will exclusively focus on $\theta=\pi$, where one can infer more on the dynamics based on the 't Hooft anomaly.

As mentioned above, an immediate consequence of the 't Hooft anomaly \eqref{anomSU2} 
for Yang-Mills theory
at $\theta=\pi$ is that the low energy theory can not be trivially gapped. What should the low energy  theory be at the fixed point?  \cite{2017JHEP...05..091G, WanWangZheng190400994} discussed  several scenarios, which we enumerate below. 
\begin{enumerate}
	\item The theory confines, and correspondingly the one-form symmetry $\Z_{2,[1]}$ is unbroken. Time reversal symmetry is spontaneously broken. There are two vacua which are related by the spontaneously broken time reversal transformation. This scenario is believed to take place for $\SU(N)$ Yang-Mills with large $N$. 
	\item The theory is gapless and deconfined, and correspondingly the one-form symmetry $\Z_{2,[1]}$ is spontaneously broken. Time reversal is unbroken. The deconfinement can be realized by a gapless conformal field theory (CFT) (e.g. U(1) Maxwell theory). See \cite{Wang2016cto1505.03520, Guo:2017xex} for discussions of different time reversal enriched gapless CFTs.  
	\item The theory is gapped and deconfined, and correspondingly the one-form symmetry $\Z_{2,[1]}$ is spontaneously broken. Time reversal is unbroken. The deconfinement can be realized by a gapped TQFT (e.g. $\Z_2$ gauge theory). In \cite{WanWangZheng190400994}, the authors have proposed the action of $\Z_2$ gauge theory in 4d saturating the anomaly \eqref{anomSU2}. 
	\item Both $\Z_{2,[1]}$ and time reversal are preserved by a gapped TQFT.  In \cite{Wan2018zql1812.11968, WanWangZheng190400994}, the authors constructed a $H$-symmetry extended TQFT via the exact sequence $1\to K\to H\to \Z_{2,[1]}\to 1$,  generalizing \cite{Wang2017locWWW1705.06728, 2018PTEP1801.05416} to higher form symmetries. By dynamically gauging $K$, it was realized that $\Z_{2,[1]}$ is spontaneously broken. This suggests a possible no go to construct a symmetric TQFT. 
	More systematically,  this scenario is ruled out by a no-go theorem from Cordova and Ohmori\cite{COnogoTQFT}, by making use of the quantum surgery constraints on cutting and gluing the spacetime manifolds\cite{Wang:2016qhf, Wang:2019diz} and other criteria.
	\item Both $\Z_{2,[1]}$ and time reversal are preserved by a gapless  CFT.
\end{enumerate} 
Though the candidate phases have been proposed, it is worthy to discuss in further detail the following aspects. 
\begin{enumerate}
	\item \cite{2017JHEP...05..091G} only discussed one sibling, i.e. $K_1=K_2=0$ among the Four Siblings in  \cite{WanWangZheng190400994}. Thus it is worthwhile to explore further the dynamical consequences for different siblings. 
	\item In the first scenario, time reversal is spontaneously broken, and there are two vacua related by time reversal symmetry.  Hence there can be a domain wall interpolating between the two vacua. The anomaly of 4d Yang-Mills \eqref{anomSU2} induces an anomaly for the 3d domain wall, hence there must be nontrivial degrees of freedom supported on the domain wall to saturate the induced anomaly.   It should be interesting to see how the four siblings of the  domain wall theory are related to each other. This will be discussed in section \ref{Sec. TimeRevDW}. 
	\item The second scenario is particularly interesting. If this scenario takes place in dynamics, the non-Abelian SU(2) gauge theory with matter can access a direct second order quantum phase transition between a U(1) spin liquid and the trivial vacuum, or more exotically between two $\U(1)$ spin liquids, depending on the further details which we discuss  in section \ref{Sec.app2}. This exotic scenario tremendously enlarges the range of possible candidates of phase transitions, and hence the multi-universality class,  between the above phases. 
\end{enumerate}
Since the $\Z_2$ gauge theory has already been studied in \cite{WanWangZheng190400994}, we will not study it in detail in the present work. We also have little to say about the last scenario.

\section{Time Reversal Domain Wall}
\label{Sec. TimeRevDW}

We consider the scenario where time reversal symmetry is spontaneously broken in the low energy. There are two vacua which are time reversal partners. Furthermore, there exists a domain wall interpolating the two vacua. Since the two time reversal breaking vacua are separately trivially gapped, the notion of domain wall theory is well defined.  The anomaly \eqref{anomSU2} implies that the domain wall theory itself has nontrivial anomaly, which enforces that the domain wall supports nontrivial degrees of freedom.

\subsection{Domain Wall for $(K_1, K_2)=(0,0)$: Semion with $\CU^2=1$}
\label{sec.DW00}

In this  subsection, we discuss the domain wall theory for the sibling $(K_1,K_2)=(0,0)$. The anomaly \eqref{anomSU2} reduces to 
$\pi\int_{M_5} B \Sq^1 B + \Sq^2 \Sq^1 B$. Under time reversal transformation, the above anomaly implies that the partition function transforms as $\bZ\to \bZ \exp(\ii \pi\int_{M_4} \CP(B)/2)$, hence induces an anomaly for the time reversal domain wall, 
\begin{eqnarray}\label{anomDW00}
S_{\mathrm{DWanom}}= \frac{\pi}{2}\int_{M_4} \CP(B).
\end{eqnarray}
The domain wall theory saturating the anomaly \eqref{anomDW00} was proposed in \cite{2017JHEP...05..091G} to be a $\SU(2)_{1}$
 Chern Simons (CS) theory 
with an action
\begin{eqnarray}\label{SU2CS}
\SU(2)_{1} \text{ CS}: ~~~~~S_{\text{CS}}=\frac{1}{4 \pi} \int_{M_3} \Tr\left(\tilde{a}  \dd \tilde{a} - \frac{2 i}{3}\tilde{a}^3\right),
\end{eqnarray}
where
$\tilde{a}$ is a one-form $\mathrm{SU}(2)$ gauge field. 
The theory \eqref{SU2CS} is a non-spin theory. There are two lines: an identity line $1$, and a line with semionic topological spin $ s$, i.e. $\{1,  s\}$. These lines obey the Abelian fusion rule: $1\times 1=1, 1\times  s =  s,  s\times  s= 1$. Hence the theory is an Abelian semion theory.  Coincidentally $\SU(2)_{1}$ is equivalent to  U(1)$_{2}$ Chern Simons. \footnote{This should be contrasted to the level rank duality $\SU(2)_{1}\longleftrightarrow U(1)_{-2}$ which only holds when both sides are regarded as spin TQFTs. }

What is the origin of the deconfined topological line $ s$ on the domain wall? We follow the discussions in \cite{Hsin2018vcg1812.04716}. Since $ s$ is also the  $\SU(2)$ Wilson line in  fundamental representation, it is natural to identify $ s$ with the $\SU(2)$ Wilson line in the fundamental representation in the 4d Yang-Mills theory, i.e., $W_{1/2} \leftrightarrow  s$. The subscript $1/2$ represents the $\SU(2)$ isospin.  However, $W_{1/2}$ in the Yang-Mills obeys area law, in accordance with the confinement. $ s$ on the domain wall has perimeter law, in accordance with the deconfinement on the wall. The behaviors of the $\SU(2)$ line in the bulk and on the wall can be understood from the different condensates in the two vacua of the bulk\cite{2017JHEP...05..091G, AharonyASY2013hda1305.0318, 1994NuPhB.426...19S, Hsin2018vcg1812.04716}.  In one vacua, confinement is due to monopole condensation. In the other vacua, confinement is due to dyon condensation. Thus although both vacua are trivially gapped, they differ by a $\Z_{2,[1]}$ symmetry protected topological (SPT) phase which is precisely described by \eqref{anomDW00}. When $W_{1/2}$ tunnels from one vacuum to the other vacuum, due to the condensate changes, $W_{1/2}$ has to deconfine on the wall. The phenomena of deconfinement can also occur on the boundary of a confining (e.g. SPT) or deconfining (e.g. SET) bulk in various dimensions, see Sec.7 of \cite{2018PTEP1801.05416} for further discussions. 

$ s$ also descends from the $\Z_{2,[1]}$ generator $U$ in 4d. The $U$ is a surface operator. In the vacuum where monopole condenses, $U$ is the spacetime trajectory of the 't Hooft line, which does not carry one-form symmetry charge itself. In the vacuum where dyons condense, the $\Z_{2,[1]}$ generator is the spacetime trajectory of the dyon line, which carries one-form symmetry charge. We consider a stretched $\Z_{2,[1]}$ generator which extends to both vacua and crosses the domain wall. $U$ and the domain wall intersects on a line, which carries $\Z_{2,[1]}$ charge which is identified as $s$. Thus $U|_{\mathrm{DW}}\leftrightarrow  s$.  Notice that semions in 3d see each other as mutual fermions. Because $ s \leftrightarrow W_{1/2}\leftrightarrow U|_{\mathrm{DW}}$, the mutual fermionic statistics between $ s$ descends from the mutual semionic statistics between $W_{1/2}$ and $U$:  $\langle W_{1/2} (\gamma) U(\Sigma)\rangle = (-1)^{\langle \Sigma, \gamma\rangle}$.

We further discuss the global symmetries of the domain wall theory $\SU(2)_{1}$. There is a $\Z_{2,[1]}$ one form global symmetry, generated by $ s$. Coupling to the background $B$ leads to the anomaly \eqref{anomDW00}. 

What about the time reversal symmetry? In the bulk, time reversal is spontaneously broken, hence time reversal exchanges the two vacua on the two sides of the domain wall. Hence time reversal is not a symmetry of the domain wall theory. In particular, time reversal $\T$ acts as 
\begin{eqnarray}
\T[\SU(2)_{1} \text{ CS}]= \SU(2)_{-1} \text{ CS}.
\end{eqnarray}
The reversed sign of the Chern Simons level reflects the reversal of the direction of the anomaly inflow under $\T$. A useful observation\cite{DWCPT1, DWCPT2} is that $\T$ can be modified to be the symmetry of $\SU(2)_{1}$ by multiplying an unbreakable $\CPoT$ in 4d. (Analogue phenomenon and more general relation to the Smith Isomorphism have been discussed by Hason, Komargodski and Thorngren \cite{DWCPT1} and independently by Cordova, Ohmori, Shao and Yan \cite{DWCPT2}. See also the talk\cite{RyanTalk} by Thorngren.  We apply this general idea to the special context: the domain wall of $\SU(2)$ Yang-Mills.) We define
\begin{eqnarray}
\CU= \T (\CPoT),
\end{eqnarray}
where $\CP_{\perp}$ is the reflection along the direction perpendicular to the domain wall. Both $\T$ and $\CPoT$ are not symmetries of $\SU(2)_{1}$, but their combination $\CU$ is. Since both $\T$ and $\CPoT$ are anti-unitary, $\CU$ is unitary.

How does $\CU$ act on the line operators in $\SU(2)_{1}$? Because both $\T$ and $\CPoT$ flip the topological spin of anyons, $\CU$ preserves the spin. Hence $\CU$ does not permute the lines. However, similar to the quantum Hall physics where anyons can transform projectively under $\U(1)$ charge conservation symmetry,  anyons can transform projectively under $\CU$. The symmetry fractionalization is classified by 
\begin{eqnarray}\label{H2}
\rH^2_{\rho}(\Z_2, \{1,  s\})= \Z_2,
\end{eqnarray}
where $\rho=1$ is the identity because $\Z_2$ symmetry generated by $\CU$ does not permute the anyons. 
To determine the action of $\CU$, we first compute $\CU^2$. Using the algebra of $\T$ and $\CPoT$ in the 4d $(K_1, K_2)=(0,0)$ Yang-Mills theory, \footnote{$\T^2=1$ is because the Wilson line is Kramers singlet.  The third equality follows from $\T (\CC \CP_{\perp})= (\CC \CP_{\perp}) \T$ which holds when acting on a bosonic line. If acting on a fermionic line,  the third equality should be modified to $\T (\CC \CP_{\perp})= -(\CC \CP_{\perp}) \T$. See section \ref{sec.CPTsummary} for further details. }
\begin{eqnarray}\label{CPTalgebra00}
\T^2=1, ~~~~ (\CPoT)^2=1, ~~~~~ \T \CPoT = \CPoT \T.
\end{eqnarray}
Thus 
\begin{eqnarray}
\CU^2= \T \CPoT \T \CPoT = \T^2 (\CPoT)^2= 1.
\end{eqnarray}
Hence $\CU$ generates a $\Z_2$ unitary symmetry that acts linearly on $W_{1/2}$. Since the Wilson line in the bulk does not transform projectively under $\T$, the Wilson line on the wall $s$ does not transform projectively under $\CU$ either.\footnote{In the next section, we will see that for the sibling $(K_1, K_2)=(1,0)$, the Wilson line transforms projectively under $\T$ and accordingly $s$ transforms projectively under $\CU$. } Thus the state $| s\rangle$ associated with the anyon $ s$ carries charge one (rather than the fractional charge) under $\CU$, i.e., $\CU | s\rangle= - | s\rangle$. In summary $\CU$ is realized linearly on the anyons which corresponds to the trivial element in \eqref{H2}. 

How does the domain wall theory couple to the  background field of $\CU$? Denote the one-form background field of $\CU$ as $Y$, satisfying $\oint Y \in 1\mod 2$. The action coupled to the background field is 
\begin{eqnarray}\label{U12Y}
\frac{2}{4\pi} \tilde{u}\dd \tilde{u}- Y \dd \tilde{u},
\end{eqnarray}
where $\tilde{u}$ is the $\U(1)$ gauge field. Here we have used the equivalence $\SU(2)_{1}\equiv U(1)_{2}$. One can check that the Wilson line $ s= \exp(\ii \oint \tilde{u})$ indeed has charge one under $\CU$. To see this, one inserts into the path integral a Wilson line along $\gamma$, which amounts to add to the action a term $\int \tilde{u} \star j$ where $\star j= \delta^{\perp}(\gamma)$. To find the $\CU$ charge of the Wilson line, we need to find the coefficient of the term $\pi Y \star j$ in the response action where the dynamical fields are integrated out. This is done by solving the equation of motion of $\tilde{u}$ and plugging back into the action \eqref{U12Y}.

Further coupling \eqref{U12Y} to $\Z_{2,[1]}$ background field $B$, the action is 
\begin{eqnarray}
\int_{M_3} \left(\frac{2}{4\pi}\tilde{u}\dd \tilde{u}- \tilde{u} B-Y\dd \tilde{u}+\pi Y B \right) +\frac{\pi}{2} \int_{M_4} \CP(B),
\end{eqnarray}
where we suppressed the cup product, e.g. $YB=Y \cup B$. 
The only anomaly is the self anomaly of $\Z_{2,[1]}$. There is no anomaly involving $\CU$.  This is also consistent with the fact that $\CU$ is not fractionalized on the anyons $\{1,  s\}$.

\subsection{Domain Wall for $(K_1, K_2)=(1,0)$: Semion with $\CU^2=-1$}
\label{sec.DW10}

We proceed to discuss the domain wall theory for the sibling $(K_1, K_2)=(1,0)$. 
Compared with the anomaly for $(K_1, K_2)=(0,0)$, the anomaly for $(K_1, K_2)=(1,0)$ contains an additional term $K_1 \pi \int w_1^2 \Sq^1 B = K_1\pi \int w_1^3 B$. Hence one may naively conclude that the anomaly for the domain wall theory is $\frac{\pi}{2} \int_{M_4} \CP(B) +  \pi \int_{M_4} w_1^2 B$. However, there are several apparent puzzles  for the above domain wall anomaly:
\begin{enumerate}
	\item Since the anomaly involves the background field $w_1$, the domain wall theory should be time reversal symmetric, and can be formulated on an unorientable manifold. However, since the 4d theory for $(K_1, K_2)=(1,0)$ only differs from $(K_1, K_2)=(0,0)$ by Lorentz symmetry fractionalization, one expects that the domain wall theory for the sibling $(1,0)$ should be a modification of $\SU(2)_{1}$ by modifying the way time reversal acts. But $\SU(2)_1$ is not time reversal symmetric in the first place and therefore does not make sense to formulate it on an unorientable manifold. 
	\item The anomaly itself, regardless of the details of the domain wall theory, is problematic. The first term $\frac{\pi}{2} \int_{M_4} \CP(B)$ is not compatible with unorientable manifold. This is because $\frac{\pi}{2} \int_{M_4} \CP(B)$  is $\Z_4$ valued, while any quantity that can be integrated on an unorientable manifold has to be $\Z_2$ valued. 
\end{enumerate}
In this section, we propose a domain wall theory by modifying the $\CU$ symmetry realization on the domain wall theory $\SU(2)_1$ proposed in section \ref{sec.DW00}, which resolves the above puzzles. 

For the sibling $(K_1, K_2)=(1,0)$, the $\SU(2)$ Wilson line in the bulk $W_{1/2}$ is a Kramers doublet, hence
\begin{eqnarray}\label{T22j}
\T^2= (-1)^{2j},
\end{eqnarray}
where $j$ is the $\SU(2)$ isospin. For our purposes, we still regard time reversal symmetry in 4d as a $\Z_2^\T$ symmetry, and \eqref{T22j} is interpreted as the Wilson line transforms in the projective representation of $\Z_2^\T$ symmetry. The algebra between $\T, \CPoT$ is (see section \ref{sec.CPTsummary} for further details)
\begin{eqnarray}\label{CPTalgebra10}
\T^2= (-1)^{2j}, ~~~(\CPoT)^2=1,~~~ \T \CPoT = \CPoT \T.
\end{eqnarray}
Hence 
\begin{eqnarray}\label{U22j}
\CU^2=\T^2= (-1)^{2j}.
\end{eqnarray}
Similar to the discussion below \eqref{T22j}, we still interpret $\CU$ as a $\Z_2$ unitary symmetry, and \eqref{U22j} implies that the anyon $s$ transforms projectively under $\CU$. Such a projective representation is the nontrivial element in \eqref{H2}.

The domain wall theory is thus $\SU(2)_{1}$ with $\Z_{2,[1]}$ one-form symmetry and $\Z_2$ zero-form symmetry generated by $\CU$, satisfying \eqref{U22j}. How does $\SU(2)_1$ couple to $\CU$ background field? As in section \ref{sec.DW00}, we still denote the $\CU$ background as $Y$ satisfying $\oint Y\in 1\mod 2$. The action coupled to the background field is 
\begin{eqnarray}\label{U1Y}
\frac{2}{4\pi} \tilde{u}\dd \tilde{u} -\frac{1}{2} Y\dd \tilde{u}.
\end{eqnarray}
Using the method discussed below \eqref{U12Y}, we find that the semion $s = \exp(\ii \oint \tilde{u})$ carries $\CU$ charge $1/2$, i.e.,  $\CU |s\rangle = \ii |s\rangle$. $\CU$ is fractionalized on $s$ as expected. 

Is the $\Z_2$ symmetry generated by $\CU$ anomalous? First it does not have anomaly with itself. To see this, we examine that under the background gauge transformation $Y\to Y+ \delta y$, \eqref{U1Y} transforms by $- \delta y \dd \tilde{u}/2$, which vanishes modulo $2\pi$. We further check the mixed anomaly between $\CU$ and $\Z_{2,[1]}$. The mixed anomaly is most conveniently seen by activating the $\Z_{2,[1]}$ background field $B$, 
\begin{eqnarray}\label{U1YB}
\int_{M_3} \left(\frac{2}{4\pi}\tilde{u}\dd \tilde{u}- \tilde{u} B-\frac{1}{2}Y \dd \tilde{u} \right) +\int_{M_4} \left(\frac{\pi}{2} \CP(B) +\pi Y Y B\right).
\end{eqnarray}
Indeed, we find two types of anomaly: $\frac{\pi}{2} \CP(B)$ is the anomaly already appeared in the domain wall for the sibling $(K_1, K_2)=(0,0)$. $\pi Y Y B$ is the mixed anomaly between $\CU$ and $\Z_{2,[1]}$, due to nontrivial $\CU$ symmetry fractionalization in \eqref{H2}. Consistently, $\pi YYB$ implies that on the domain wall, the $\Z_{2,[1]}$ generator $s$ is attached by a surface operator $\exp(\ii \pi \int_{\Sigma} YY)=\exp(\ii \pi/2 \int_{\Sigma} \delta Y)=\exp(\ii \pi/2 \oint_{\partial \Sigma}Y)$ which precisely reflects the fact that $s$ carries $Y$ charge $1/2$. We make several comments:
\begin{enumerate}
	\item In \eqref{U1YB}, the domain wall  theory $\SU(2)_{1}$ is not time reversal symmetric. Consistently, the anomaly does not involve $w_1$, which resolves the two puzzles mentioned in the beginning of this subsection. 
	\item The time reversal symmetry fractionalization in the 4d induces a unitary $\Z_2$ symmetry fractionalization on the domain wall. Correspondingly, the mixed $\T-\Z_{2,[1]}$ anomaly $\pi w_1^3 B$ induces a mixed $\CU-\Z_{2,[1]}$ anomaly $\pi YY B$ on the domain wall. 
	\item Since $\CPoT$ is always an unbreakable symmetry in 4d Yang-Mills, one can freely modify the $\T$ background field $w_1$ to $\T (\CPoT)$ background field $Y$. Hence the anomaly $\pi w_1^3 B$ for Yang-Mills can be equivalently be written as $\pi w_1 YY B$. This rewriting makes the induced  anomaly $\pi Y Y B$ of the domain wall natural, because under time reversal, 4d Yang-Mills partition function transforms as
	\begin{eqnarray}\label{Zvar}
	\bZ\to \bZ\exp\left(\frac{\ii \pi }{2} \int_{M_4}  \CP(B) +\ii \pi  \int_{M_4} Y Y B\right),
	\end{eqnarray}
	which naturally provides the anomaly inflow of the 3d domain wall theory \eqref{U1YB}. We emphasize that in \eqref{U1YB}, one can not replace $Y$ by $w_1$ back, because $\CPoT$ is no longer the symmetry of the domain wall. 
\end{enumerate}

\subsection{Domain Wall for $(K_1, K_2)=(0,1)$: Anti-Semion with $\CU^2=1$}
\label{sec.DW01}

We proceed to discuss the domain wall theory for the sibling $(K_1, K_2)=(0,1)$. We will find that although the anomaly for 4d Yang-Mills does not depend on $K_2$, the anomaly of the domain wall does! To see this, we rewrite $K_2$ dependent term in \eqref{anomSU2} as $K_2 \pi \Sq^1(w_2 B) = K_2 \pi w_1 w_2 B$ which does not vanish on a manifold with boundary. This term induces an anomaly on the domain wall $K_2 \pi w_2 B$. The complete anomaly for the domain wall is 
\begin{eqnarray}\label{anomDW01}
S_{\mathrm{DWanom}}= \frac{\pi}{2} \int_{M_4} \CP(B) +  \pi \int_{M_4} w_2 B.
\end{eqnarray}
We look for the domain wall theory that saturates such an anomaly. 

We start with $\SU(2)_{1}\equiv \U(1)_{2}$ theory for the sibling $(K_1, K_2)=(0,0)$. We have shown in section \ref{sec.DW00} that $\SU(2)_{1}$ saturates the first term in \eqref{anomDW01}. One needs to find a proper fractionalization of the Lorentz symmetry (whose background is $w_2$) to further match the anomaly $\pi w_2 B$.   Denote the topological spin of the $\Z_{2,[1]}$ generator $ s$ in $\SU(2)_{1}$ as $h(s)$. The  additional anomaly $\pi \int w_2 B$ modifies the topological spin of  $ s$ by\cite{2019arXiv190907383H}
\begin{eqnarray}
h( s) \to h( s)+ \frac{1}{2} \mod 1.
\end{eqnarray}
Hence after symmetry fractionalization, $h(s)$ shifts from $1/4$ to $3/4$ mod 1. In other words, the semion in the domain wall for the sibling $(K_1, K_2)=(0,0)$ becomes an anti-semion for the domain wall in the sibling $(K_1, K_2)=(0,1)$. Thus the domain wall TQFT for $(K_1, K_2)=(0,1)$ contains a trivial anyon and an anti-semion, i.e. $\{1, \bar s\}$. Such a TQFT is precisely 
\begin{eqnarray}
\SU(2)_{-1}~ \mathrm{CS}.
\end{eqnarray}

Apart from using Lorentz symmetry fractionalization, the domain wall theory can further be obtained by rewriting the anomaly \eqref{anomDW01} as 
\begin{eqnarray}\label{anomDW01p}
S_{\mathrm{DWanom}}= \frac{\pi}{2} \int_{M_4} \CP(B) +  \pi \int_{M_4} \CP(B) = \frac{3\pi}{2} \int_{M_4} \CP(B) = -\frac{\pi}{2} \int_{M_4} \CP(B).
\end{eqnarray}
Comparing with the anomaly \eqref{anomDW00}, the anomaly \eqref{anomDW01p} simply changes the sign, i.e., the direction of the anomaly inflow is reversed. Consistently, the level of the  domain wall Chern Simons theory is also reversed, from $\SU(2)_1$ for the sibling $(K_1, K_2)=(0,0)$ to $\SU(2)_{-1}$ for the sibling $(K_1, K_2)=(0,1)$.

The Lorentz symmetry fractionalization can also be viewed from the quantum number of Wilson line $W_{1/2}$ in the 4d Yang-Mills. For the sibling $(K_1, K_2)=(0,1)$,  $W_{1/2}$ transforms projectively under the $\SO(3,1)$ Lorentz rotation, hence it is a fermion. As explained in section \ref{sec.DW00}, the deconfined line $\bar s$ is obtained from the Wilson line in the bulk. Hence the Lorentz symmetry fractionalization (the shift of statistics by $1/2$) for $W_{1/2}$ naturally induces a Lorentz symmetry fractionalization (the shift of statistics by $1/2$) for $s$ on the domain wall, which yields $\bar s$, consistent with the additional anomaly $\pi w_2 B$ for the domain wall.

It is instructive to consider the fractionalization of the unitary $\Z_2$ symmetry $\CU$ on $\bar s$. In the 4d Yang-Mills of the sibling $(K_1, K_2)=(0,1)$, the Wilson line $W_{1/2}$ 
is a Kramers doublet, i.e., $\T^2=-1$. More generally, $\T^2=(-1)^{2j}$ where $j$ is the $\SU(2)$ isospin. Hence using the algebra of $\T$ and $\CPoT$,  (see section \ref{sec.CPTsummary} for further details)
\begin{eqnarray}\label{CPTalgebra01}
\T^2=(-1)^{2j}, ~~~~ (\CPoT)^2=1, ~~~~~ \T (\CPoT) = (-1)^{2j}(\CPoT) \T,
\end{eqnarray}
we find 
\begin{eqnarray}
\CU^2= (-1)^{2j}\T^2= (-1)^{4j}=1.
\end{eqnarray}
It is ramarkable that although the time reversal symmetry is fractionalized on Wilson line $W_{1/2}$ in the bulk, $\CU$ is \emph{not} fractionalized on the anyon $\bar s$ !    
Hence similar to the case in section \ref{sec.DW00}, the anti-semion $\bar s$ transforms linearly under $\CU$, i.e., $\CU^2(\bar s)=\bar s$.   We further couple $\SU(2)_{-1}$ to both $\Z_{2,[1]}$ background field $B$ and the $\CU$ background field $Y$, 
\begin{eqnarray}
\int_{M_3} \left(-\frac{2}{4\pi}\tilde{u}\dd \tilde{u}+\tilde{u}B+Y\dd \tilde{u}-\pi Y B \right) + \int_{M_4} \left(\frac{\pi}{2}\CP(B)+ \pi w_2 B\right).
\end{eqnarray}
The fact that $\CU$ is not fractionalized on $\bar s$ is in accord with the fact that there is no anomaly involve $\CU$ on the wall.

\subsection{Domain Wall for $(K_1, K_2)=(1,1)$: Anti-Semion with $\CU^2=-1$}

We finally discuss the domain wall theory for the sibling $(K_1, K_2)=(1,1)$. From the discussion in section \ref{sec.DW10} and \ref{sec.DW01}, we find that the anomaly for the domain wall theory is
\begin{eqnarray}
S_{\mathrm{DWanom}}= \frac{\pi}{2} \int_{M_4} \CP(B) +  \pi \int_{M_4} (Y Y+ w_2) B,
\end{eqnarray}
where $Y$ is the background field for the unitary symmetry $\CU=\T (\CPoT)$. The domain wall theory is $\SU(2)_{-1}$ properly coupled to background fields $Y$ and $w_2$:
\begin{eqnarray}
\int_{M_3} \left(-\frac{2}{4\pi}\tilde{u}\dd \tilde{u}+ \tilde{u}B+\frac{1}{2}Y \dd \tilde{u} \right) +\int_{M_4} \left(\frac{\pi}{2} \CP(B)+ \pi w_2 B +\pi Y Y B\right).
\end{eqnarray}

We emphasize that although  time reversal is not fractionalized on the $W_{1/2}$ in the bulk, i.e., $\T^2=1$, the $\CU$ unitary symmetry is fractionalized on the anyon $\bar s$.  Furthermore, we again observe that domain wall carries nontrivial anomaly related to $w_2$, although the bulk does not.

\subsection{Remarks On $\CC\CP_{\perp}$ and $\T$, and Summary}
\label{sec.CPTsummary}

We provide some additional remarks on the 4d symmetries $\CC\CP_{\perp}$ and $\T$. The purpose is to further explain the algebra between $\T$ and $\CPoT$, i.e. \eqref{CPTalgebra00}, \eqref{CPTalgebra10} and \eqref{CPTalgebra01}. 

As mentioned in section \ref{sec.YM}, for the sibling $(K_1, K_2)$, the Wilson line $W_{1/2}$ has spin $h(W)= K_2/2$, which is explained below \eqref{GBC3}.  However, the fact that time reversal squares to be $\T^2= (-1)^{K_1+K_2}$, rather than $\T^2= (-1)^{K_1}$, needs further explanation, which we provide below. (See \cite{Hsin2019fhf1904.11550} for similar explanation in 4d Maxwell theory. ) 
For $K_2=0$, $\T^2=(\CC\CP_{\perp})^2 = (-1)^{K_1+K_2}= (-1)^{K_1}$, hence $K_1=0,1$ represents Kramers singlet and doublet respectively. However for $K_2=1$, suppose when we move from a Minkowski spacetime to a Euclidean spacetime, $\T$ becomes a Euclidean reflection $\CR$ via a Wick rotation. Then $\T^2$ differs by a sign from $\CR^2$, i.e. $\T^2=-\CR^2$. Such a minus sign only occurs when acting on a fermion. Notice that in Minkowski spacetime,  $\CC\CP_{\perp}$ is a still a Euclidean reflection, so $\T^2= - (\CC\CP_{\perp})^2$. To synthesize, we have 
\begin{eqnarray}
\T^2= (-1)^{K_2} (\CC\CP_{\perp})^2.
\end{eqnarray}
If $\T^2= (-1)^{K_1+K_2}$, then $(\CC\CP_{\perp})^2=(-1)^{K_1}$, hence 
\begin{eqnarray}
\T (\CC\CP_{\perp})  =  \CPoT \CPoT \T (\CC\CP_{\perp}) =(\CPoT) \T^2 (\CC\CP_{\perp})^2= \CPoT (-1)^{K_2},
\end{eqnarray}
where we used $(\CPoT)^2=1$. This further gives rise to the commutation relation between $\T$ and $\CPoT$ as
\begin{eqnarray}
\T (\CPoT) = (-1)^{K_2} (\CPoT) \T.
\end{eqnarray}
This is precisely the relation in \eqref{CPTalgebra00}, \eqref{CPTalgebra10} and \eqref{CPTalgebra01}.

We summarize the symmetry properties of the Wilson lines of isospin $j$ in the bulk, the domain wall theory, their symmetry fractinoalization pattern and the anomalies in table \ref{Tab:DW4sib}.
\begin{table}[H]
		\centering
		\setstretch{1.8}
		\begin{tabular}{ |c|c| c| c | c| } 
			\hline
        $(K_1, K_2)$ & $(h\mod 1, \T^2)$ & DW Theory & $\CU^2$ & DW Anomaly\\
        \hline
        $(0,0)$ & $(0,1)$ & $\SU(2)_1=\{1,s\}$ & $1$ & $\frac{\pi}{2}\int_{M_4} \CP(B)$\\
        \hline
        $(1,0)$ & $(0,(-1)^{2j})$ & $\SU(2)_1=\{1,s\}$ & $(-1)^{2j}$ & $\frac{\pi}{2}\int_{M_4} \CP(B)+ \pi \int_{M_4} YY B$\\
        \hline
        $(0,1)$ & $(j,(-1)^{2j})$ & $\SU(2)_{-1}=\{1,\bar s\}$ & $1$ & $\frac{\pi}{2}\int_{M_4} \CP(B)+ \pi \int_{M_4} w_2 B$\\
        \hline 
        $(1,1)$ & $(j,1)$ & $\SU(2)_{-1}=\{1,\bar s\}$ & $(-1)^{2j}$ & $\frac{\pi}{2}\int_{M_4} \CP(B)+ \pi \int_{M_4} (YY+w_2)B$\\
        \hline
		\end{tabular}
		\caption{Symmetry fractionalization and anomalies on the domain wall theory for four siblings of Yang-Mills. }
		\label{Tab:DW4sib}
\end{table}

\section{Application I: Domain Wall Theory $N_f<N_{CFT}$}
\label{sec.app1}

We start by considering the domain wall theory for $\SU(2)$ QCD with $N_f$ fermions. The theory depends on the mass and the theta parameter via $m^{N_f}e^{\ii\theta}$. In this section, we assume $m$ to be real and non-negative, and keep $\theta$ in the Lagrangian. (In section \ref{Sec.app2}, we will adopt the different assumption.)  We exclusively focus on $\theta=\pi$, which is time reversal symmetric. We denote $\Lambda$ as the strong coupling scale. 

When $m \gg \Lambda$, one can integrate out the massive fermions, and the low energy effective theory is the $\SU(2)$ Yang-Mills theory with $\theta=\pi$. Assuming the scenario where the time reversal is spontaneously broken, there are two vacua which are time reversal partners. Between the two vacua, there is a time reversal domain wall. We further assume that $N_f$ is below the conformal window, i.e. $N_f<N_{CFT}$, the domain wall theory has been conjectured to be\cite{2018JHEP...01..110G}
\begin{eqnarray}\label{duality}
\SU(2)_{1-\frac{N_f}{2}} + N_f \psi \longleftrightarrow \U(1)_{-2} + N_f \phi.
\end{eqnarray}

In the large mass limit, the domain wall theory \eqref{duality} flows to $\SU(2)_1\equiv \U(1)_2$, corresponding to the domain wall theory of the pure $\SU(2)$ Yang-Mills at $\theta=\pi$. As discussed in section \ref{Sec. TimeRevDW}, there are multiple versions of $\SU(2)_1$ theories, distinguished by the enrichments of the unitary symmetry $\CU$ and the Lorentz symmetry. In this section, we determine the symmetry enriched versions of $\SU(2)_{1-\frac{N_f}{2}}+ N_f \psi$, and how the symmetry enrichments match across the duality \eqref{duality}.

\subsection{Lorentz Symmetry Fractionalization, $K_2=1$}
\label{sec.k2=1}

We first show that domain wall theory realized in $\SU(2)$ QCD requires $K_2=1$. In the bulk, since the $\SU(2)$ gauge field is coupled to fermions, the $2\pi$ Lorentz rotation, which multiplies the fermions by $-1$, can be compensated by a $\SU(2)$ gauge transformation. More precisely, the gauge-spacetime symmetry is 
\begin{eqnarray}
\frac{\SU(2)\times \Spin(3,1)}{\Z_2},
\end{eqnarray}
and the constraint  of the symmetry bundle  is 
\begin{eqnarray}
w_2(V_{\SO(3)})= w_2.
\end{eqnarray}
Comparing with \eqref{GBC3}, we find that the Lorentz symmetry $\SO(3,1)$ is always realized projectively, hence the effective Yang-Mills corresponds to the sibling $K_2=1$. As discussed in section \ref{sec.DW01}, in the large mass limit on the domain wall $\SU(2)_1$, the $\SU(2)$ gauge bundle in the domain wall theory is also twisted by the Lorentz symmetry SO(2,1), i.e. the gauge-spacetime symmetry on the domain wall, as well as the bundle constraint are 
\begin{eqnarray}\label{DWbundle}
\mathrm{Domain~Wall}: ~~~~~\frac{\SU(2)\times \Spin(2,1)}{\Z_2}, ~~~~~ w_2(V_{\SO(3)})= w_2.
\end{eqnarray}
Thus at large mass limit on the wall, the Chern Simons is the $K_2=1$ enrichment of $\SU(2)_1$, i.e. $\SU(2)_{-1}$ Chern Simons theory discussed in section \ref{sec.DW01}. Notice that this is precisely the large positive mass limit on the bosonic side of \eqref{duality}. Hence the $\SO(2,1)$ Lorentz symmetry fractionalization is matched across the duality on the wall. See \cite{2019arXiv190907383H} for more examples.

\subsection{$\CU$ Unitary Symmetry Fractionalization}

We proceed to discuss the fractionalization of $\Z_2$ unitary symmetry generated by $\CU$ on the domain wall. We first consider the large positive mass limit in the theory $\SU(2)_{1-N_f/2}+N_f \psi$. There are two options of fractionalization of $\CU$ on the anti-semion $\bar s$,\footnote{Notice that Lorentz symmetry fractionalization of the semion results in an anti-semion. } labeled by $K_1$. Concretely, there is the correspondence
\begin{eqnarray}
\CU^2=(-1)^{K_1} ~\text{on~anti-semion}~\bar s.
\end{eqnarray}
When the mass of $\psi$ is finite, the $\Z_2$ unitary symmetry acts on the fermion $\psi$. For $K_1=0$, the fermion carries charge $1$, while for $K_1=1$, the fermion carries charge $1/2$ (i.e. fractionalized). 

On the other hand, notice that $\SU(2)_{1-N_f/2}+N_f \psi$ naturally has the $\U(1)$ symmetry associated with Baryon conservation, and we adopt the normalization that the Baryon has $\U(1)$ charge $2$, while the quark $\psi$ has $\U(1)$ charge $1$. The symmetry is 
\begin{eqnarray}
\frac{\U(1)\times \SU(2)\times \Spin(2,1)}{\Z_2\times \Z_2},
\end{eqnarray} 
where the constraint between the bundles is $w_2(V_{\SO(3)})+ c_1(V_{\U(1)/\Z_2})+ w_2=0\mod 2$.

How does the $\Z_2$ symmetry generated by $\CU$ relate to $\U(1)$?  For $K_1=0$, the quark $\psi$ carries $\CU$ charge 1, hence the $\Z_2$ is embedded in $\U(1)$ in the natural way, i.e.  $\Z_2\subset \U(1)$. For $K_1=1$, the quark $\psi$ carries $\CU$ charge $1/2$, hence the $\Z_2$ is embedded into $\U(1)$ as 
$\Z_2\subset \U(1)/\Z_2$, or equivalently $\Z_4\subset \U(1)$. We enumerate the total $\CU$-gauge-spacetime symmetry and their gauge bundle constraint as follows:
\begin{equation}\label{GBC3d}
\begin{split}
(K_1,K_2)=(0,1): &~~~~~ \frac{\Z_2\times \SU(2)\times \Spin(2,1)}{\Z_2\times \Z_2}, ~~~~~ w_2(V_{SO(3)})+w_2=0\mod 2,\\
(K_1,K_2)=(1,1): &~~~~~ \frac{\Z_4\times \SU(2)\times \Spin(2,1)}{\Z_2\times \Z_2}, ~~~~~\Sq^1Y+ w_2(V_{SO(3)})+w_2=0\mod 2.\\
\end{split}
\end{equation}
Notice that the gauge bundle constraints for the domain wall theories \eqref{GBC3d} are nicely in accord with \eqref{GBC3} in 4d.  

Let us consider the dual theory $\U(1)_{-2}+ N_f\phi$, and discuss how the $\CU$ symmetry is realized. We first consider the large mass limit, where the theory flows to $\U(1)_{-2}$. The monopole in the bosonic theory is dual to the Baryon in the fermionic theory. In the fermionic theory, Baryon carries $\U(1)$ charge $2$. Using the embedding of $\Z_2$ into $\U(1)$, we find that Baryon carries $\CU$ charge $K_1\mod 2$.  Thus the monopole carries $\CU$ charge $K_1\mod 2$. 

The symmetry breaking quantum phase (described by the nonlinear sigma model) on the domain wall can be easily seen from the bosonic theory. By turning on the large negative mass squared of the scalar, we land on the symmetry breaking phase described by the nonlinear sigma model with the target space
\begin{eqnarray}
G= \frac{\Sp(2)}{\Sp(1)\times \Sp(1)}= \frac{\Sp(2)}{\Spin(4)}.
\end{eqnarray}
In the sigma model, there exists a configuration of skyrmion which also carries the $\CU$ charge $K_1\mod 2$.

\section{Deconfined Gapless U(1) Gauge Theory}
\label{Sec.U1}

In this section, we discuss the scenario where the low energy theory of $\SU(2)$ Yang-Mills at $\theta=\pi$ is a $\U(1)$ gauge theory.\footnote{We will also comment on $\theta=0$. } We attempt to find a $\U(1)$ gauge theory that matches the anomaly \eqref{anomSU2}.

We consider the time reversal invariant $\U(1)$ gauge theory described by the action 
\begin{eqnarray}\label{U(1)gaugetheory}
S= -\frac{1}{4e^2} \int_{M_4} f \wedge \star f + \frac{\theta}{8\pi^2} \int_{M_4} f\wedge f, ~~~~~ \theta=0, 2\pi,
\end{eqnarray}
where $f= \dd u$ and $u$ is the $\U(1)$ gauge field. The $\U(1)$ theory is time reversal symmetric, where $\Z_2^\T$ acts on the gauge field as 
\begin{eqnarray}
\T(u_{0}(t, \vec x))=-u_{0}(-t, \vec x), ~~~~~ \T(u_{i}(t, \vec x))=u_{i}(-t, \vec x).
\end{eqnarray}
This choice of time reversal flips the $\U(1)$ gauge charge, while preserves the $\U(1)$ gauge monopole. Hence one can assign monopole Kramers degeneracy, i.e., $\T^2$ to the lines with charge $(q_e, q_m)=(0, 1)$.\footnote{For $\theta=0$, the dyonic line with charge $(q_e, q_m)=(0, 1)$ is denoted the 't Hooft line. However, for $\theta=2\pi$, due to Witten effect, the 't Hooft line $T$ is has charge $(q_e, q_m)=(1,1)$. The dyonic line with charge $(q_e, q_m)=(0, 1)$ is $W^{-1}T$, i.e.  't Hooft line attached with an anti-Wilson line. } In the present case, $\T^2=1$ acting on Wilson lines. 
Under Lorentz rotation, the Wilson line transforms with integer spin, while the 't Hooft line transforms with half integer spin or integer spin depending on $\theta=0, 2\pi$, due to the statistical Witten effect.

\eqref{U(1)gaugetheory} also has one form symmetries $\U(1)_{e,[1]}\times \U(1)_{m, [1]}$ where the subscripts $e$ and $m$ represent electric and magnetic respectively. The electric $\U(1)_{e,[1]}$ acts on Wilson lines, and $\U(1)_{m,[1]}$ acts on 't Hooft lines. To make contact with the $\SU(2)$ Yang-Mills,  we will focus on the $\Z_{2, [1]}$ subgroup of $\U(1)_{e,[1]}$. To couple \eqref{U(1)gaugetheory} to two-form background gauge field $B$, we replace $f$ by $f-\pi B$. The action is 
\begin{eqnarray}
S= -\frac{1}{4e^2} \int_{M_4} (f-\pi B) \wedge \star (f-\pi B) + \frac{\theta}{8\pi^2} \int_{M_4} (f-\pi B)\wedge (f-\pi B), ~~~\theta=0, 2\pi.
\end{eqnarray}
We further discuss coupling \eqref{U(1)gaugetheory} to the Lorentz background fields $w_1, w_2$.

\subsection{$\U(1)$ Gauge Theory and Spin Liquids at $\theta=0$}
\label{sec.U1theta0}

For $\theta=0$, one can further couple \eqref{U(1)gaugetheory} to the Lorentz background fields.  Changing $B\to B+J_2 w_2$ modifies the statistics of the $\U(1)$ charge. To modify the Lorentz symmetries of the $\U(1)$ monopole, we add to the action a term 
\begin{eqnarray}
\frac{1}{2} (f- \pi B-  J_2 \pi w_2) (L_1 w_1^2+ L_2 w_2).
\end{eqnarray}
The Lorentz quantum numbers of the $\U(1)$ charge $\tilde{E}$ and the $\U(1)$ monopole $\tilde{M}$ are
\begin{eqnarray}
\begin{split}
\tilde{E}: &~~~~~ h(\tilde{E})= \frac{J_2}{2}\mod 1\\
\tilde{M}:& ~~~~~ h(\tilde{M})= \frac{L_2}{2}\mod 1, ~~~~~ \T^2_{\tilde{M}}= (-1)^{L_1+L_2},
\end{split}
\end{eqnarray}
where we use the tilde to emphasize that the time reversal parities of the $\U(1)$ charge and monopole are the opposite compared with the convention in \cite{Wang2016cto1505.03520}, namely the time reversal flips the charge $\tilde{E}$ other than the monopole $\tilde{M}$. We will bridge  both conventions at the end of this section.

When coupled to all the background fields $B, w_1, w_2$ (i.e. by formulating the theory on an unorientable and non-spin manifold), the $\U(1)$ gauge theory with $\theta=0$ is
\begin{equation}
S=-\frac{1}{4e^2}\int_{M_4} (f-\pi B - J_2 \pi w_2)\wedge \star (f-\pi B - J_2 \pi w_2) + \frac{1}{2} \int_{M_4}  (f- \pi B-  J_2 \pi w_2) \wedge (L_1 w_1^2+ L_2 w_2).
\end{equation}
The last term $-\frac{1}{2} \int_{M_4}(\pi B+ J_2 \pi w_2) \wedge (L_1 w_1^2+ L_2 w_2)\subset S$ is not well-defined on an unorientable manifold. To make sense of it on an unorientable manifold, we need to promote it to a 5d action, 
\begin{eqnarray}\label{theta0anom}
-\pi \int_{M_5}\Sq^1 \left((B+ J_2 w_2) (L_1 w_1^2+ L_2 w_2)\right).
\end{eqnarray}
Among the four terms by expanding \eqref{theta0anom}, only two terms represent the 't Hooft anomalies,  
\begin{eqnarray}
S_{\mathrm{anom}}= -\pi \int_{M_5} (L_1 w_1^2 \Sq^1 B+ J_2 L_2 w_2 w_3),
\end{eqnarray} 
where $w_3\equiv w_3(TM_5)$ is the Stiefel-Whitney class for the tangent bundle of $M_5$. When $L_1=1$, there is a mixed anomaly between the time reversal and $\Z_{2,[1]}$. When $J_2=L_2=1$, there is an anomaly for the ``all fermion electrodynamics"\cite{Kravec:2014aza, Wang2016cto1505.03520, WWW1810}.

We summarize the $\U(1)$ gauge theories at $\theta=0$ and their 't Hooft anomalies as
\begin{eqnarray}\label{U1SLtheta0}
\begin{split}
(J_2, L_2, L_1)=(0,0,0)& ~~~~~ \tilde{E}_b\tilde{M}_b~~~~~ ~~0,\\
(J_2, L_2, L_1)=(0,0,1)& ~~~~~ \tilde{E}_b\tilde{M}_{bT}~~~~~ -\pi \int_{M_5}  w_1^2 \Sq^1 B,\\
(J_2, L_2, L_1)=(0,1,0)& ~~~~~ \tilde{E}_b\tilde{M}_{fT}~~~~~0,\\
(J_2, L_2, L_1)=(0,1,1)& ~~~~~ \tilde{E}_b\tilde{M}_{f}~~~~~~ -\pi \int_{M_5}  w_1^2 \Sq^1 B,\\
(J_2, L_2, L_1)=(1,0,0)& ~~~~~ \tilde{E}_f\tilde{M}_b~~~~~~~0,\\
(J_2, L_2, L_1)=(1,0,1)& ~~~~~ \tilde{E}_f\tilde{M}_{bT}~~~~ -\pi \int_{M_5}  w_1^2 \Sq^1 B,\\
(J_2, L_2, L_1)=(1,1,0)& ~~~~~ \tilde{E}_f\tilde{M}_{fT}~~~~ -\pi \int_{M_5} w_2w_3,\\
(J_2, L_2, L_1)=(1,1,1)& ~~~~~ \tilde{E}_f\tilde{M}_{f}~~~~~~ -\pi \int_{M_5}  w_1^2 \Sq^1 B+ w_2 w_3,\\
\end{split}
\end{eqnarray}
where we used the Lorentz symmetries of the $\U(1)$ charge and $\U(1)$ monopoles to label the spin liquid, similar to \cite{Wang2016cto1505.03520}. However, we emphasize that $\tilde{E}$ is time reversal odd and $\tilde{M}$ is time reversal even, in contrast to the conventions of \cite{Wang2016cto1505.03520} where the time reversal parities are the opposite to ours. 

Comparing with the anomalies of $\SU(2)$ Yang-Mills with $\theta=\pi$ \eqref{anomSU2}, none of the $\U(1)$ spin liquids in \eqref{U1SLtheta0} can be the potential IR candidate phases of $\SU(2)$ Yang-Mills at $\theta=\pi$. However, we will see in section \ref{Sec.app2} that some of the $\U(1)$ spin liquids in \eqref{U1SLtheta0} can be obtained by Higgsing $\SU(2)$ gauge group to $\U(1)$ for the $\SU(2)$ Yang-Mills with $\theta=0$, although it is very \emph{unlikely} that the deconfined $\U(1)$ spin liquids are dynamically realized by the RG flow.

It is illuminating to connect our identification of the $\U(1)$ spin liquids to those in \cite{Wang2016cto1505.03520}. In \cite{Wang2016cto1505.03520}, the convention is that $\U(1)$ charge $E$ is time reversal even while the $\U(1)$ monopole $M$ is time reversal odd. For $\theta=0$, the two conventions are related by $S$-duality, i.e. $E\leftrightarrow \tilde{M}, M\leftrightarrow \tilde{E}$ which can be understood as the $\pi/2$ rotation of the charge-monopole lattice.  Thus we arrive at the following dictionary:
\begin{eqnarray}\label{corr}
\begin{split}
&\tilde{E}_b\tilde{M}_b= E_bM_b, ~~~\tilde{E}_b\tilde{M}_{bT}= E_{bT}M_{b}, ~~~ \tilde{E}_b\tilde{M}_{fT}= E_{fT} M_{b}, ~~~ \tilde{E}_b\tilde{M}_{f}= E_{f} M_{b},\\
&\tilde{E}_f\tilde{M}_b=E_bM_f, ~~~\tilde{E}_f\tilde{M}_{bT}= E_{bT}M_{f}, ~~~\tilde{E}_f\tilde{M}_{fT}= E_{fT}M_{f}, ~~~\tilde{E}_f\tilde{M}_{f}= E_{f}M_{f}.
\end{split}
\end{eqnarray}
In the dual theory, only $M$ is charged under $\Z_{2, [1]}$.

\subsection{$\U(1)$ Gauge theory and Spin Liquids at $\theta=2\pi$ }
\label{sec.U1theta2pi}

We proceed to discuss the $\U(1)$ spin liquids with $\theta=2\pi$. Similar to section \ref{sec.U1theta0}, one can still modify the statistics of the $\U(1)$ charge  by replacing $B\to B+ J_2 w_2$. To modify the Lorentz symmetries of the monopole with charge $(q_e, q_m)=(0,1)$, we realize that due to the Witten effect, the 't Hooft operator ('t Hooft line) carries $\theta/2\pi=1$ electric charge. Thus to form the pure monopole with vanishing electric charge, one needs to attach a $\U(1)$ charge (i.e. a Wilson line). As noted in \cite{Hsin2019fhf1904.11550}, for $\theta=2\pi$, a dyon with charge $(q_e, q_m)$ couples to the $\U(1)_{e,[1]}$ and $\U(1)_{m,[1]}$ background fields $B_e$ and  $B_m$ by attaching a surface operator
\begin{eqnarray}\label{surface}
\exp\left( \ii  \int_{\Sigma} (q_e-q_m)B_e + q_m B_m + (q_e-q_m) q_m \pi w_2 \right).
\end{eqnarray}
Applying \eqref{surface} to our case, $B_e= \pi (B+ J_2 w_2)$.  We demand that when $B=0$, the surface operator for $(q_e, q_m)=(0,1)$ should be $L_1 w_1^2+ L_2 w_2$. As we will see below, to match the mixed anomaly between time reversal and $\Z_{2, [1]}$, we need to modify the above expression to $L_1 w_1^2+ L_2 w_2+B$ when $B$ is nonvanishing. This implies that both $\tilde{E}$ and $\tilde{M}$ are charged under $\Z_{2, [1]}$, and the mixed $\T$-$\Z_{2, [1]}$ anomaly descends from the mixed anomaly of $\Z_{2, [1]}\subset \U(1)_{e,[1]}$ and $\Z_{2, [1]}\subset \U(1)_{m,[1]}$. 
The Lorentz symmetry of the $\U(1)$ charge $\tilde{E}$ and $\U(1)$ monopole $\tilde{M}$ is 
\begin{eqnarray}\label{EMsym2pi}
\begin{split}
\tilde{E}: &~~~~~ h(\tilde{E})= \frac{J_2}{2}\mod 1\\
\tilde{M}:& ~~~~~ h(\tilde{M})= \frac{L_2}{2}\mod 1, ~~~~~ \T^2_{\tilde{M}}= (-1)^{L_1+L_2}.
\end{split}
\end{eqnarray}
Thus we find 
\begin{eqnarray}\label{BeBm}
B_e= \pi (B+ J_2 w_2), ~~~~~ B_m= \pi \left(L_1 w_1^2 + (L_2+J_2+1)w_2\right).
\end{eqnarray}
Notice that the Yang-Mills couples to $\U(1)_{e,[1]}$ and $\U(1)_{m,[1]}$ background fields $B_e$ and $B_m$ as 
\begin{equation}\label{U1bgd}
S= -\frac{1}{4e^2} \int_{M_4} (f-B_e) \wedge \star (f-B_e) + \frac{2\pi}{8\pi^2} \int_{M_4} (f-B_e)\wedge (f-B_e)+ \frac{\pi}{2\pi} \int_{M_4} (f-B_e)B_m.
\end{equation}
Substituting \eqref{BeBm} into \eqref{U1bgd}, we obtain the $\U(1)$ gauge theory coupled to $B, w_1, w_2$ as
\begin{eqnarray}\label{U12pi}
\begin{split}
S=& -\frac{1}{4e^2} \int_{M_4} (f-\pi B- J_2\pi w_2 ) \wedge \star  (f-\pi B- J_2\pi w_2 ) \\&+ \frac{2\pi}{8\pi^2} \int_{M_4}  (f-\pi B- J_2\pi w_2 )\wedge (f-\pi B- J_2\pi w_2 )\\&+ \frac{\pi}{2\pi} \int_{M_4}  (f-\pi B- J_2\pi w_2 )\left(L_1 w_1^2 + (L_2+J_2+1)w_2\right).
\end{split}
\end{eqnarray}
The anomaly can be derived by examining the terms in \eqref{U12pi} that are not well-defined on an unorientable manifold $M_4$. Such terms are $\frac{2\pi}{8\pi^2}\int_{M_4} (\pi B+ J_2 \pi w_2)^2 - \frac{\pi}{2\pi}\int_{M_4}(\pi B+ J_2 \pi w_2)(L_1 w_1^2 + (L_2 +J_2+1)w_2) $ due to the fractional coefficients. To make sense of these terms, we promote these terms to a 5d integral.
\begin{equation}
S_{\mathrm{anom}}= \pi \int_{M_5} B \Sq^1 B+ \Sq^2 \Sq^1 B + (L_2+1) \Sq^1(w_2 B)+ L_1 w_1^2 \Sq^1 B + J_2(L_2+J_2+1)w_2 w_3.
\end{equation}
As commented in section \ref{sec.SU2YM}, the term $ \Sq^1(w_2 B)$ is a WZW-like counter term.

We summarize the $\U(1)$ spin liquids with $\theta=2\pi$ and their genuine 't Hooft anomalies (i.e. excluding the WZW-like counter terms) as follows:
\begin{equation}\label{U1SLtheta2pi}
\begin{split}
(J_2, L_2, L_1)=(0,0,0)& ~~~~~ (\tilde{E}_b\tilde{M}_b)_{2\pi}~~~~~ \pi\int_{M_5} B \Sq^1 B+ \Sq^2 \Sq^1 B,\\
(J_2, L_2, L_1)=(0,0,1)& ~~~~~ (\tilde{E}_b\tilde{M}_{bT})_{2\pi}~~~~~ \pi \int_{M_5} B \Sq^1 B+ \Sq^2 \Sq^1 B+ w_1^2 \Sq^1 B,\\
(J_2, L_2, L_1)=(0,1,0)& ~~~~~ (\tilde{E}_b\tilde{M}_{fT})_{2\pi}~~~~~ \pi \int_{M_5} B \Sq^1 B+ \Sq^2 \Sq^1 B,\\
(J_2, L_2, L_1)=(0,1,1)& ~~~~~ (\tilde{E}_b\tilde{M}_{f})_{2\pi}~~~~~~ \pi \int_{M_5} B \Sq^1 B+ \Sq^2 \Sq^1 B+ w_1^2 \Sq^1 B,\\
(J_2, L_2, L_1)=(1,0,0)& ~~~~~ (\tilde{E}_f\tilde{M}_b)_{2\pi}~~~~~\pi \int_{M_5} B \Sq^1 B+ \Sq^2 \Sq^1 B,\\
(J_2, L_2, L_1)=(1,0,1)& ~~~~~ (\tilde{E}_f\tilde{M}_{bT})_{2\pi}~~~~ \pi \int_{M_5} B \Sq^1 B+ \Sq^2 \Sq^1 B+  w_1^2 \Sq^1 B,\\
(J_2, L_2, L_1)=(1,1,0)& ~~~~~ (\tilde{E}_f\tilde{M}_{fT})_{2\pi}~~~~ \pi \int_{M_5}B \Sq^1 B+ \Sq^2 \Sq^1 B+ w_2w_3,\\
(J_2, L_2, L_1)=(1,1,1)& ~~~~~( \tilde{E}_f\tilde{M}_{f})_{2\pi}~~~~~~ \pi \int_{M_5} \Sq^1 B+ \Sq^2 \Sq^1 B+ w_1^2 \Sq^1 B+ w_2 w_3.\\
\end{split}
\end{equation}
We use the subscript $2\pi$ to emphasize that both $\tilde{E}$ and $\tilde{M}$ lines are charged under $\Z_{2,[1]}$.  
By rotating the charge-monopole lattice by $\pi/2$ (i.e. performing the $S$-duality), we are also able to map the $\U(1)$ spin liquids in \eqref{U1SLtheta2pi} to those discussed in \cite{Wang2016cto1505.03520}. One simply exchange $\tilde{E}\leftrightarrow M$ and $\tilde{M}\leftrightarrow E$. The correspondence has been enumerated in \eqref{corr}. We emphasize that the $\Z_{2, [1]}$ one form symmetry background field couples to both $E$ and $M$ lines in the dual theory. 

Notice the WZW-like counter term does not have to be matched along the RG flow. By matching the genuine 't Hooft anomalies in \eqref{U1SLtheta2pi} and the anomalies of $\SU(2)$ Yang-Mills at $\theta=\pi$, we can enumerate the $\U(1)$ spin liquids for each sibling of $\SU(2)$ Yang-Mills as follows:  
\begin{eqnarray}\label{match}
\begin{split}
(K_1, K_2)=(0,0), (0,1): &~~~~~ (\tilde{E}_b\tilde{M}_b)_{2\pi}, ~ (\tilde{E}_b\tilde{M}_{fT})_{2\pi},~ (\tilde{E}_f\tilde{M}_b)_{2\pi}.\\
(K_1, K_2)=(1,0), (1,1): &~~~~~ (\tilde{E}_b\tilde{M}_{bT})_{2\pi}, ~(\tilde{E}_b\tilde{M}_{f})_{2\pi}, ~(\tilde{E}_f\tilde{M}_{bT})_{2\pi}.
\end{split}
\end{eqnarray}
The remaining two $\U(1)$ spin liquids can not emerge under the RG flow of any sibling of $\SU(2)$ Yang-Mills due to the additional $w_2 w_3$ anomaly. Merely from matching the 't Hooft anomalies, we are not able to determine which among the three $\U(1)$ spin liquids in each row of \eqref{match} is realized for a given $(K_1, K_2)$. However, by imposing more physical requirements as we will discuss in section \ref{Sec.app2}, we are able to determine which $\U(1)$ spin liquid phase is realized.

\section{Application II: Gauge Enhanced Quantum Critical Point  $N_f\geq N_{CFT}$}
\label{Sec.app2}

In this section, we discuss an application of the deconfinement scenario in section \ref{Sec.U1}. Assuming the $\SU(2)$ Yang-Mills at $\theta=\pi$ can flow to a deconfined $\U(1)$ gauge theory which describes the low energy physics of the $\U(1)$ quantum spin liquid, it opens up the possibility of exotic quantum phase transitions between different $\U(1)$ spin liquids and/or trivial paramagnet, where the gauge group is enhanced to $\SU(2)$ at and only at the critical point. We denote such transition as a gauge enhanced quantum critical point (GEQCP).

\subsection{$\SU(2)$ QCD$_4$ and Higher Order Interactions: $\U(1)$ Spin Liquid Phases From Higgsing }
\label{SU2QCD}

We consider  the $\SU(2)$ QCD$_4$ with $N_f$ fermions, described by the following action 
\begin{equation}\label{eq:QCD+interaction}
S= \int_{M_4}\left( \sum_{i=1}^{N_f} \bar{\Psi}_i(\ii \gamma^\mu D_\mu-m)\Psi_i + \CL_{\mathrm{high}}\right)-\frac{1}{4g^2}\int_{M_4}\Tr(F \wedge \star F).
\end{equation}
where $\Psi_i$ is the four component Dirac fermion with a flavor index $i=1, ..., N_f$ and a $ \SU(2)$ color index $a=1,2$ which is suppressed.  For the sake of the following discussion,  we have also included a phenomenological  four and eight-fermion interaction term $\CL_{\mathrm{high}}$, 
\begin{eqnarray}\label{high}
\CL_{\mathrm{high}}=u\sum_{a=1}^{3}\left(\sum_{i=1}^{N_f}\bar{\Psi}_i\tau^a\Psi_i\right)^2 + \lambda \left[\sum_{a=1}^{3}\left(\sum_{i=1}^{N_f}\bar{\Psi}_i\tau^a\Psi_i\right)^2 \right]^2,
\end{eqnarray}
where $\tau^a$ ($a=1,2,3$) denotes the generator of the $\SU(2)$ gauge group. We will always take $\lambda>0$ and allow $u$ to be either sign. Throughout, we assume there is a flavor symmetry $\Sp(N_f)$ or $\U(N_f)$ such that the masses of all the flavors of fermions are degenerate.

\begin{figure}[htbp]
	\begin{center}
		\includegraphics[width=0.5\columnwidth]{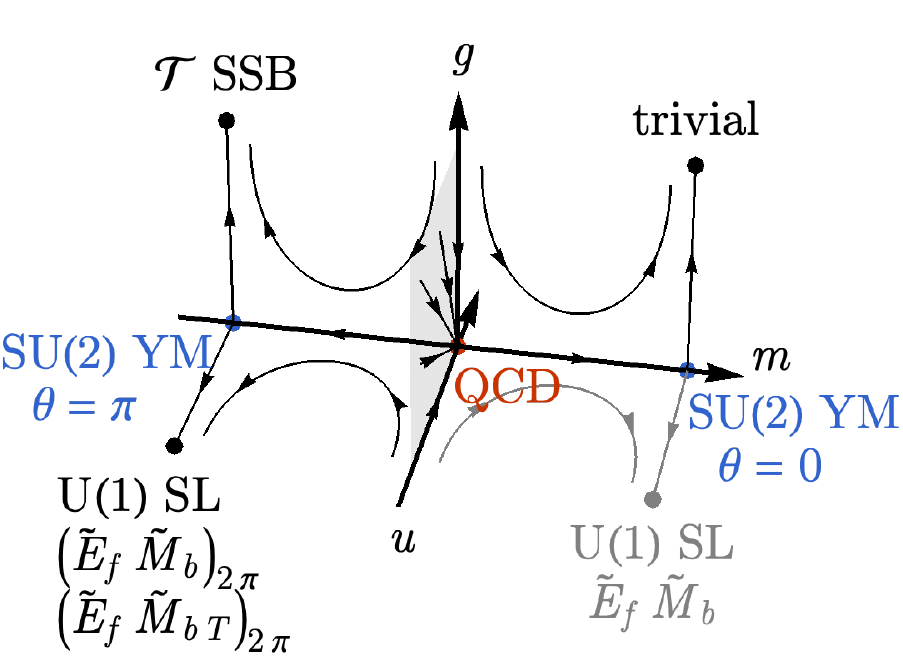}
		\caption{Schematic RG flow diagram around the QCD$_4$ fixed point for odd $N_f$ and $N_f>11$. Possible IR fates are listed for completeness, although some (such as the U(1) SL on the $\theta=0$ side) may be extremely  unlikely.}
		\label{fig:RGflow}
	\end{center}
\end{figure}

We work in the parameter regime of $N_f\geq N_{CFT}$ such that the QCD$_4$ with $m=0$ flows to a conformal field theory which can describe a second order phase transition between the two semi-classical phases (which we will discuss in detail below). In particular, when $N_f> 11$, the QCD$_4$ with $m=0$ is in the infrared free phase and the coupling constant $g$ flows to zero under RG.  At this RG fixed point, the only relevant perturbation is the fermion mass $m$, and the terms in $\CL_{\mathrm{high}}$ are irrelevant. Thus for $m=0$, adding the higher order terms $\CL_{\mathrm{high}}$ in \eqref{eq:QCD+interaction} does not affect the dynamics in the IR. In particular, $u, g$ and $\lambda$ all flow to zero, as shown in the middle panel of figure \ref{fig:RGflow}.

We proceed to discuss the mass deformation by allowing $m$ to be either positive or negative.\footnote{In general, the mass parameter in 4d QCD can be complex, which is obvious when we rewrite the Dirac fermions into Weyl fermions with both chirality. In this work, we focus on the real mass for simplicity.} We focus on the case when $N_f$ is an odd integer. Then depending on the sign of $m$, the QCD  flows to the $\SU(2)$ Yang-Mills theory with either $\theta=0$ (for $m>0$) or $\theta=\pi$ (for $m<0$). The $\SU(2)$ Yang-Mills theory does not describe the ultimate IR fate of the system. It continues to flow towards different possible IR fixed points as we have discussed in previous sections. One possibility is that the system enters the confinement phase, where the coupling $g$ flows large away from the  $m=0$ QCD$_4$ fixed point. In the confinement phase on the $\theta=\pi$ ($m<0$) side, $\Z_{2,[1]}$ is unbroken and time reversal symmetry is spontaneously broken \cite{2017JHEP...05..091G, TSBDQCP}.  Another possibility is that the system remains deconfined with a reduced gauge group, which can lead to either a $\U(1)$ or a $\mathbb{Z}_2$ spin liquid phase. The possible $\U(1)$ spin liquid phases that saturate the 't Hooft anomalies are provided in section \ref{Sec.U1}. In the rest of this section, we provide a potentially possible mechanism for the deconfinement scenario to take place, and we further determine, if so,  which type of $\U(1)$ spin liquid (among the candidates in \eqref{match})  is indeed realized for a given sibling of $\SU(2)$ Yang-Mills. 

Viewing the $\SU(2)$ Yang-Mills as the large mass deformation limit of a $\SU(2)$ QCD$_4$ allows us to propose a natural mechanism to realize the deconfinement scenario in section \ref{Sec.U1}.  When $|m|$ is nonzero, it is possible that the interaction strength $u$ and $\lambda$ in \eqref{eq:QCD+interaction} and \eqref{high} could flow strong. Assuming $u<0$,  the higher order term \eqref{high} drives the condensation of $\SU(2)$ gauge triplet $\bar{\Psi}_i\bm{\tau}\Psi_i$ and consequently Higgses the $\SU(2)$ gauge group to its subgroup. If only one component of the gauge triplet acquires expectation value, e.g.~$\langle\bar{\Psi}_i\tau^3\Psi_i\rangle\neq0$, the $\SU(2)$ gauge group will be Higgsed down to its $\U(1)$ subgroup. The remaining low-energy theory will be a $\U(1)$ Maxwell theory that describes the $\U(1)$ spin liquid. It will be important in section \ref{sec.SymfracU1} that after Higgsing, each flavor of $\Psi_i$ gives rise to  two types of fermions $\Psi_{1i}, \Psi_{2i}$ which carry opposite $\U(1)$ gauge charge. $\Psi_{1i}$ carries $\U(1)$ charge $1$, while $\Psi_{2i}$ carries $\U(1)$ charge $-1$.  If more than one components of the gauge triplet acquire expectation values (depending on the details of higher order interactions), e.g.~$\langle\bar{\Psi}_i\tau^1\Psi_i\rangle,\langle\bar{\Psi}_i\tau^2\Psi_i\rangle\neq0$, then the remaining gauge group will be $\mathbb{Z}_2$, realizing the TQFT description of the topologically ordered $\mathbb{Z}_2$ spin liquid phase.(See \cite{WanWangZheng190400994} for such $\Z_2$ spin liquid phases.) In the following, we will take the $\U(1)$ spin liquid as the example to illustrate the deconfined phase. The schematic RG flow diagram is shown in figure \ref{fig:RGflow}.

We comment on the possibilities of the signs of $m$ and $u$ in \eqref{eq:QCD+interaction} and \eqref{high}, and their consequences. 
\begin{enumerate}
	\item \textbf{$u>0$ for both $m>0$ and $m<0$}: In this scenario, the gauge group $\SU(2)$ is not Higgsed. When $m$ is positive, the theory flows to a trivial gapped phase, in accord with the standard lore.\cite{YMMP-Jaffe-Witten} When $m$ is negative, the theory flows to a strongly coupled confining phase where time reversal is spontaneously broken. 
	\item \textbf{$u<0$ for both $m>0$ and $m<0$}: In this scenario, the gauge group $\SU(2)$ is Higgsed for both signs of $m$, with the only exception at $m=0$. The $\SU(2)$ Yang-Mills with both $\theta=0$ and $\pi$ flow to certain $\U(1)$ spin liquids. We will determine on the $\U(1)$ spin liquid in section \ref{sec.SymfracU1}. We emphasize that although it is extremely unlikely that $\SU(2)$ Yang-Mills with $\theta=0$ flows to a deconfined $U(1)$ gauge theory and is beyond the standard lore \cite{YMMP-Jaffe-Witten}, this scenario is still not completely ruled out rigorously. 
	\item \textbf{$u>0$ for $m>0$,  and $u<0$ for $m<0$}: The signs of $m$ and $u$ are correlated. In this scenario, the gauge group $\SU(2)$ is Higgsed only for $\theta=\pi$. While for $\theta=0$, the $\SU(2)$ Yang-Mills flows to a trivial gapped phase, consistent with the lore \cite{YMMP-Jaffe-Witten}. However, the underlying mechanism for the sign correlation between $u$ and $m$ still needs to be understood. 
\end{enumerate}

\subsection{Symmetries Realizations and Symmetry Enriched $\U(1)$ Spin Liquids in the Infrared}
\label{sec.SymfracU1}

The specific type of the $\U(1)$ spin liquid that is realized under the gauge triplet condensation depends on how the time-reversal symmetry is implemented in the QCD theory \eqref{eq:QCD+interaction}. We consider the following two possibilities of time reversal implementation, where the gauge and global symmetries are 
\begin{eqnarray}
&\mathrm{CI}:& \frac{\SU(2)\times \Sp(N_f)\times\mathbb{Z}_4^\mathcal{T}}{\mathbb{Z}_2^c\times\mathbb{Z}_2^f},\label{CI}\\ 
&\mathrm{CII}:& \frac{\SU(2)\times \U(N_f)}{\mathbb{Z}_2^f}\times\mathbb{Z}_2^\mathcal{T}. \label{CII}
\end{eqnarray}

For \eqref{CI}, the $\SU(2)\equiv\Sp(1)$ gauge transformation and the time reversal symmetry act on the fermionic matter field as $\SU(2):\Psi_i \to e^{\ii \bm{\theta}\cdot\bm{\tau}}\Psi_i$ and
\begin{eqnarray}\label{CITR}
\mathrm{CI}: ~~~~~\mathcal{T}: \Psi_i \to \mathcal{K}\gamma^5 \gamma^0\Psi_i^\dagger.
\end{eqnarray}
In particular $\T^2=-1$ on $\Psi_i$. 
Here the $\mathbb{Z}_2^c$ center of $\SU(2)$
is the same as the fermion parity $\mathbb{Z}_2^f$; we mod out
$\mathbb{Z}_2^c=\mathbb{Z}_2^f$ twice because
${\SU(2)}$, ${\Sp(N_f)}$ and ${\mathbb{Z}_4^\mathcal{T}}$
all share the same normal subgroup $\mathbb{Z}_2^c=\mathbb{Z}_2^f$. $\Sp(N_f)$ is the flavor symmetry.  If we just focus on the $\SU(2)$ and time reversal (i.e. ignore the flavor symmetry $\Sp(N_f)$), this symmetry coincides with the $\mathrm{CI}$ symmetry class in the ten fold classification of the fermionic SPT. This motivates an alternative way to understand the $\SU(2)$ QCD$_4$ \eqref{eq:QCD+interaction}: The $\SU(2)$ QCD$_4$ with symmetry class \eqref{CI} can be understood as from gauging the $\SU(2)$ global symmetries of $N_f$ copies of  free fermions in symmetry class CI.

For \eqref{CII}, the $\SU(2)$ gauge transformation acts in the same way as in the CI class. However, time reversal  acts on the fermionic matter field differently:
\begin{eqnarray}\label{CIITR}
\mathrm{CII}: ~~~~~\T: \Psi_{i} \to \mathcal{K}\ii \gamma^5 \gamma^0  \Psi_{i}^\dagger.
\end{eqnarray}
Compared with \eqref{CITR}, there is an additional $\U(1)\subset \U(N_f)$ flavor transformation. 
In particular, $\T^2=1$. The quotient in \eqref{CII} is to identify the common normal subgroup of $\SU(2)$ and $\U(N_f)$. The $\SU(2)$ QCD$_4$ with symmetry class \eqref{CII} can be understood as from gauging the $\SU(2)$ global symmetries of $N_f$ copies of free fermions in symmetry class CII.

Under the condensation of $\langle\bar{\Psi}_i\tau^3\Psi_i\rangle\neq0$, the remaining $\U(1)$ gauge group acts as $\U(1):\Psi_i\to e^{\ii \theta\tau^3}\Psi_i$. The $\U(1)$ generator commutes with the time reversal transformation, which forms the AIII symmetry class.  The class AIII fermionic SPT state is $\mathbb{Z}_{8}\times \Z_2$ classified, where only the phases associated with $\Z_8$ can be represented  by the free fermion theories.\footnote{Before gauging, the AIII SPT theory is simply $N_f$ free fermions coupled to $\U(1)$ background fields. Hence only the $\Z_8$ part is relevant for our purpose. } 
Turning on the fermion mass $m$ effectively put the $\Psi_i$ field in the class AIII fermionic SPT states labeled by the topological index $\nu=0$ ($m>0$) or $\nu=2N_f$ ($m<0$). Connecting with the $\U(1)$ gauge theories in section \ref{Sec.U1}, $\theta=\nu \pi$.  If $\Psi_i$ is in the $\nu=0$ phase, the $\U(1)$ monopole is simply a boson. If $\Psi_i$ is in the $\nu=2N_f$ phase, the $\U(1)$ monopole will carry will carry $2N_f$ fermion zero modes. However, these zero modes carry $\U(1)$ gauge charge. To form $\U(1)$ gauge invariant monopole operator, we need to consider only those monopole that are neutral under $\U(1)$.  We note that under Higgsing, both CI and CII classes reduce to AIII classes, 
\begin{eqnarray}\label{higgsing}
\text{Higgsing}: ~~~~~ \mathrm{CI}\to \mathrm{AIII}, ~~~~~ \mathrm{CII}\to \mathrm{AIII}.
\end{eqnarray}
\eqref{higgsing} can also be interpreted as different ways of embedding AIII symmetry class into CI and CII classes. See \cite{Guo:2017xex} for extensive discussions of the embedding in \eqref{higgsing} and other examples among the ten Cartan symmetry classes.   The difference between the two reduced $\mathrm{AIII}$ classes are that the $\U(1)$ neutral monopoles have different symmetry quantum numbers, which we determine below.

We proceed to determine the time reversal properties (Kramers degeneracy) of the time reversal symmetric monopole operators of charge $(q_e, q_m)=(0,1)$. For illustrative purposes, 
we first determine the time reversal properties of the monopole in AIII class $\nu=2$ (i.e. $N_f=1$ copy of AIII system and the topological theta parameter in the $\U(1)$ Mexwell theory is $\theta=2\pi$) with the global symmetry
\begin{equation}
\frac{\U(1)\times \Z_4^\T}{\Z_2^f}.
\end{equation}
The time reversal properties of the fermion zero modes descend from the time reversal transformations in \eqref{CITR} and \eqref{CIITR}. In \eqref{CITR}, time reversal maps a fermion to its conjugate, and only the spinor indices are rotated. Hence the fermion zero mode $c_{a}$ (for $N_f=1$), where $a=1, 2$ is the $\SU(2)$ index, maps under time reversal as 
\begin{eqnarray}\label{CITRc}
\mathrm{CI}: ~~~~~ \T: ~~~c_{a}\to c_{a}^\dagger,~~ c_a^\dagger\to c_a.
\end{eqnarray}
In \eqref{CIITR}, time reversal maps a fermion to its conjugate, accompanied by a $\Z_4\subset \U(1)$ transformation generated by $\ii$. Hence the fermion zero mode $c_{a}$ maps under time reversal as 
\begin{eqnarray}\label{CIITRc}
\mathrm{CII}: ~~~~~  \T: ~~~c_{a}\to \ii c_{a}^\dagger,~~c_{a}^\dagger\to -\ii c_a.
\end{eqnarray}
Using the operator-state correspondence, the monopole $\CM$ without any fermion zero mode occupied is mapped to a state $|0\rangle$ with $c_a|0\rangle=0$ for $a=1,2$. Under $\T$, the empty state $|0\rangle$ is mapped to a fully occupied state $c_1^\dagger c_2^\dagger|0\rangle$, i.e. $\T|0\rangle=c_1^\dagger c_2^\dagger|0\rangle$.
We further notice that the two fermion zero modes has opposite gauge charge. $c_1$ carries $\U(1)$ charge 1, while $c_2$ carries $\U(1)$ charge $-1$. Thus the $\U(1)$ neutral monopole operators are associated with the states
\begin{eqnarray}\label{emptyfull}
|0\rangle, ~~~ c_1^\dagger c_2^\dagger|0\rangle
\end{eqnarray}
rather than the half filled states $c_1^\dagger|0\rangle, c_2^\dagger |0\rangle$. 
Combined with \eqref{CITRc} and \eqref{CIITRc}, we can compute $\T^2$ of the empty and full states in \eqref{emptyfull} as 
\begin{equation}\label{TRmono}
\begin{split}
\mathrm{CI}:& ~~~~~ \T^2 |0\rangle =  c_1 c_2 c_1^\dagger c_2^\dagger|0\rangle= - |0\rangle, ~~~~~ \T^2 c_1^\dagger c_2^\dagger|0\rangle = c_1^\dagger c_2^\dagger c_1 c_2 c_1^\dagger c_2^\dagger|0\rangle= -c_1^\dagger c_2^\dagger |0\rangle,\\
\mathrm{CII}: &~~~~~ \T^2 |0\rangle =  -c_1 c_2 c_1^\dagger c_2^\dagger|0\rangle=  |0\rangle, ~~~~~ \T^2 c_1^\dagger c_2^\dagger|0\rangle  = -c_1^\dagger c_2^\dagger c_1 c_2 c_1^\dagger c_2^\dagger|0\rangle= c_1^\dagger c_2^\dagger |0\rangle.
\end{split}
\end{equation}
In short, for $N_f=1$ (or $\nu=2$), the $(q_e, q_m)=(0,1)$ monopole is Kramers doublet ($\T^2=-1$) in the AIII class descended from CI, while Kramers singlet ($\T^2=1$) in the AIII class descended from CII. Moreover, in both cases, the $(q_e, q_m)=(0,1)$ monopole is a boson, which is obvious from \eqref{EMsym2pi}. 

Using similar analysis for the monopole quantum numbers in \eqref{TRmono} for the AIII class $\nu=2$, it is straightforward to obtain the monopole quantum numbers for AIII class $\nu=2N_f$, which is
\begin{eqnarray}\label{mono}
\begin{split}
\mathrm{CI}:&~~~~~ \T^2=(-1)^{N_f},\\
\mathrm{CII}:&~~~~~ \T^2=1,
\end{split}
\end{eqnarray}
for monopoles associated with the $\U(1)$ neutral states $|0\rangle, c_{1i}^\dagger c_{2 j}^\dagger|0\rangle, ..., (c_{1 i_1}^\dagger...c_{1 i_{N_f}}^\dagger c_{2 j_{1}}^\dagger ... c_{2 j_{N_f}}^\dagger)|0\rangle$ where the number of $1$ and $2$ of the $\SU(2)$ indices should balance. Since we focus on the case where $N_f$ is odd, the time reversal Kramers degeneracy for the two cases in \eqref{mono} are different.

We emphasize that the  quantum numbers in \eqref{mono} are for the probe monopoles in the AIII symmetry classes. Further gauging the $\U(1)$ global symmetries of the AIII fermionic SPTs lead to different $\U(1)$ spin liquids. Thus \eqref{mono} also characterizes the quantum numbers of the dynamical monopoles in the $\U(1)$ spin liquids.

We are ready to identify the $\U(1)$ spin liquid phases in the IR. We first determine the candidate $\U(1)$ spin liquid for $\SU(2)$ Yang-Mills with $\theta=0$.  Since in $\SU(2)$ QCD$_4$, the $\SU(2)$ gauge field is coupled to fermions, the $\SU(2)$ Yang-Mills theories should have fermionic Wilson lines, i.e., $K_2=1$. (See an similar discussion in section \ref{sec.k2=1}. ) On the other hand, the $\U(1)$ charges should also be fermionic because they descend from Higgsing the fermionic $\SU(2)$ charges, i.e. $\tilde{E}$ should be fermionic. Combining with the $\U(1)$ monopole quantum numbers in \eqref{mono}, we find that, when $m<0$,  the QCD in the CI class flows to $(\tilde{E}_f \tilde{M}_{bT})_{2\pi}$, while the QCD in CII class flows to $(\tilde{E}_f \tilde{M}_{b})_{2\pi}$. 

The make contact with the siblings of $\SU(2)$ Yang-Mills at $\theta=\pi$, we further need relate the $\U(1)$ spin liquids determined above to the labels of the siblings, i.e. $(K_1, K_2)$. As we find above, $K_2=1$. Furthermore, $K_1$ can be determined by matching the anomaly of the $\U(1)$ spin liquids in \eqref{U1SLtheta2pi} with the anomaly of the $\SU(2)$ Yang-Mills \eqref{anomSU2}. Thus we 
we determine the $\U(1)$ spin liquids as well as the siblings of the Lorentz symmetry enriched $\SU(2)$ Yang-Mills as
\begin{eqnarray}
\nu=2N_f: ~~~&\mathrm{CI}:& (K_1, K_2)=(1,1), ~~~~~ \to ~~~~~ \mathrm{AIII}: (\tilde{E}_f \tilde{M}_{bT})_{2\pi}\label{CI2N}.\\
\nu=2N_f: ~~~&\mathrm{CII}:& (K_1, K_2)=(0,1), ~~~~~ \to ~~~~~ \mathrm{AIII}: (\tilde{E}_f \tilde{M}_{b})_{2\pi}\label{CII2N}.
\end{eqnarray}
The $\U(1)$ spin liquids on the $m>0$ side is simply 
\begin{eqnarray}\label{CICII0}
\nu=0:~~~ \mathrm{CI}, \mathrm{CII}:(K_1, K_1)=(0,1),(1,1)~~~~~\to ~~~~~\mathrm{AIII}:\tilde{E}_f \tilde{M}_b.
\end{eqnarray}
Thus we have singled out a particular symmetry enriched $\U(1)$ spin liquid as the low energy of $\SU(2)$ Yang-Mills from the anomaly matched candidates in \eqref{match}, by embedding the $\SU(2)$ Yang-Mills into a $\SU(2)$ QCD$_4$ with the assumed $\SU(2)$ triplet Higgsing pattern. \eqref{CI2N} and \eqref{CII2N} are precisely the time reversal CFTs initially proposed \cite{Guo:2017xex}.

We finally comment that although $\Psi_i$ in \eqref{CITR} satisfies $\T^2=-1$, this does not mean $\Psi_i$ is Kramers doublet, because $\Psi_i$ is not mapped to itself under time reversal. See \cite{2014Sci...343..629W} for an analogue discussion in $2+1$d. A priori, it seems to be difficult to determine the $(K_1, K_2)$ from the symmetry assignment \eqref{CITR}. Here, we provide a way to determine it through identifying the $\U(1)$ spin liquid $(\tilde{E}_f \tilde{M}_{bT})_{2\pi}$ and via anomaly matching. Analogue comments also apply to \eqref{CIITR}.

\subsection{Gauge Enhanced Quantum Critical Points}

From the $\U(1)$ spin liquids determined in section \ref{sec.SymfracU1}, we are able to predict a series of gauge enhanced quantum critical points (GEQCP) using $\SU(2)$ QCD$_4$. We will focus on the second and third scenarios in section \ref{SU2QCD} which involve $\U(1)$ spin liquid phases, and finally comment on the first scenario where no $\U(1)$ spin liquid phases are involved. 

We first discuss the second scenario in section \ref{SU2QCD} where the fermion bilinear condensation takes place for both $m>0$ and $m<0$, realizing $\tilde{E}_f\tilde{M}_b$ and $(\tilde{E}_f \tilde{M}_{bT})_{2\pi}$ respectively for the sibling $(K_1, K_2)=(1,1)$, while $\tilde{E}_f\tilde{M}_b$  and $(\tilde{E}_f \tilde{M}_{b})_{2\pi}$ respectively for the sibling $(K_1, K_2)=(0,1)$. For simplicity, we will mainly discuss the sibling $(K_1, K_2)=(1,1)$ below. 
The transition between $\tilde{E}_f\tilde{M}_b$ and $(\tilde{E}_f \tilde{M}_{bT})_{2\pi}$ spin liquids can be realized by tuning the mass $m$ in \eqref{eq:QCD+interaction}, assuming the $\SU(2)$ Yang-Mills theory can flow to the deconfined $\U(1)$ Maxwell theory on both sides. At $m=0$, both the gauge coupling $g$ and the interaction $u$ are irrelevant (if $N_f> 11$), such that the transition is controlled by the IR free QCD fixed point. This provides a novel GEQCP scenario for the Kramer-changing quantum criticality between $\tilde{E}_f\tilde{M}_b$  and $(\tilde{E}_f \tilde{M}_{bT})_{2\pi}$  spin liquids as a QCD theory, where the gauge group is enhanced from $\U(1)$ to $\SU(2)$ at the critical point, which is different from the QED description proposed in Ref.\cite{Wang2016cto1505.03520}. Nevertheless, similar to Ref.\cite{Wang2016cto1505.03520}, additional symmetries must be imposed to guarantee a single direct transition, otherwise the critical point can be interrupted by other time reversal invariant terms such as the alternating chemical potential term $\psi_i^\dagger \gamma^5\psi_i$ or can be split to multiple transitions if different fermion flavors have different masses. One simple way is to demand an inversion symmetry $\mathcal{I}:\psi_i\to\gamma_0\psi_i, \bm{x}\to-\bm{x}$ together with the $\Sp(N_f)$ flavor symmetry.

We proceed to the third scenario in in section \ref{SU2QCD} where the fermion bilinear condensation takes place only for  $m<0$, realizing $(\tilde{E}_f \tilde{M}_{bT})_{2\pi}$  for the sibling $(K_1, K_2)=(1,1)$, while  $(\tilde{E}_f \tilde{M}_{b})_{2\pi}$  for the sibling $(K_1, K_2)=(0,1)$. On the $m>0$ side, the theory flows to a trivial vacua. For simplicity, we only discuss the sibling $(K_1, K_2)=(1,1)$. 
The QCD theory also afford a GEQCP scenario for the phase transition between the $(\tilde{E}_f \tilde{M}_{bT})_{2\pi}$ $\U(1)$ gauge theory and the trivially confined vacuum. The conventional transition from a $\tilde{E}_f\tilde{M}_b$ $\U(1)$ spin liquid to a trivial paramagnet can happen by
monopole condensation (as a confinement transition). Note that $\tilde{E}$ is a fermion and can not be condensed, unless condensing in pairs which would lead to a $\Z_2$ topological order. However for the $(\tilde{E}_f \tilde{M}_{bT})_{2\pi}$ spin liquid, if we condense the monopole, the time reversal symmetry will be spontaneously broken because the monopole is a Kramers doublet.
It seems difficult to drive a direct transition from the $(\tilde{E}_f \tilde{M}_{bT})_{2\pi}$ spin liquid to a trivial paramagnet. Nevertheless, our analysis provides a compelling possibility by first enlarging the gauge group from $\U(1)$ to $\SU(2)$ and then allowing the $\SU(2)$ to confine trivially by removing tuning to the $\theta=0$ side. As shown in the flow diagram Fig.\ref{fig:RGflow}, it is possible to connect the $(\tilde{E}_f \tilde{M}_{bT})_{2\pi}$ spin liquid and the trivial paramagnet in the parameter space by going through the plane of $m=0$, which is controlled by the QCD fixed point, where an enlarged $\SU(2)$ gauge group together with gapless fermionic partons will emerge. This constitutes yet another example of the GEQCP.

We finally comment on the first scenario, where time reversal is spontaneously broken for $m<0$, and a trivial gapped phase is realized for $m>0$. If this scenario takes place, the $\SU(2)$ QCD$_4$ with odd $N_f$ fundamental fermions can access as a  second order  deconfined phase transition, where deconfinement is realized at and only at the critical point. This scenario is discussed in \cite{TSBDQCP}.  See \cite{Bi:2018xvr, 2018arXiv180609592C, AnberPoppitz2018tcj1805.12290, Wan2018djlW2.1812.11955} for other deconfined quantum critical points (DQCP) between various confining phases.

\bigskip

\emph{Note Added}: 
After the completion of this work, \cite{TSBDQCP} appeared, which partially overlaps with our work. In particular, our discussion of the more general GEQCP has been partly motivated by the talk given by T. Senthil at the Ultra Quantum Matter kickoff meeting (Sep. 12, 2019) about the possible DQCP towards the symmetry breaking phases\cite{SenthilTalk}. After the completion of our work, we learnt that similar discussion about the $\U(1)$ spin liquid and its transition to a trivial insulator  had also appeared in Sec.V of \cite{TSBDQCP}. Our results agree with \cite{TSBDQCP} when they overlap, in particular $(\tilde{E}_f \tilde{M}_{bT})_{2\pi}$ in our work is denoted as $E_{f\frac{1}{2}}M_{\frac{1}{2}T}$ in \cite{TSBDQCP}.

\section*{Acknowledgements}

The authors are listed in the alphabetical order by the standard convention.  We thank Zhen Bi, Clay Cordova,  Ho-Tat Lam, John McGreevy, Shu-Heng Shao, Chong Wang and Liujun Zou for discussions. We thank T.Senthil  for email correspondences and insightful comments on the manuscript.  
J.W. is supported by
NSF Grant PHY-1606531. 
Y.Z. thanks the support from Andrei Bernevig and from Physics Department of Princeton University.
Part of this work was done during the Workshop on Topological Aspects of Condensed Matter at CMSA. 
This work is also supported by 
NSF Grant DMS-1607871 ``Analysis, Geometry and Mathematical Physics'' 
and Center for Mathematical Sciences and Applications at Harvard University.

\bibliographystyle{utphys}
\bibliography{YMDynamics_QCP}

\end{document}